\documentclass[floats,floatfix,showpacs,amssymb,prd,twocolumn,superscriptaddress,nofootinbib,nolongbibliography,reprint]{revtex4-1}

\usepackage{amssymb,amsmath,verbatim,mathtools,needspace,enumitem,etoolbox,graphicx,physics,microtype,afterpage,xspace,tabularx,lmodern,multirow}
\usepackage{gensymb}
\usepackage[normalem]{ulem}
\usepackage[dvipsnames, usenames]{xcolor}
\definecolor{linkcolor}{rgb}{0.0,0.3,0.5}
\usepackage[unicode, colorlinks=true, linkcolor=linkcolor, citecolor=linkcolor, filecolor=linkcolor, urlcolor=linkcolor, linktocpage, breaklinks]{hyperref}
\usepackage[all]{hypcap}
\usepackage[T1]{fontenc}
\usepackage[utf8]{inputenc}
\usepackage[usenames,dvipsnames]{xcolor}
\hypersetup{colorlinks=true,citecolor=romared,linkcolor=romared,urlcolor=romared}

\definecolor{romared}{RGB}{142,0,28}

\newcommand{\be}{\begin{equation}}
\newcommand{\ee}{\end{equation}}

\def\be{\begin{equation}}
\def\ee{\end{equation}}
\newcommand{\beq}{\begin{eqnarray}}
\newcommand{\eeq}{\end{eqnarray}}

\usepackage{aas_macros}
\usepackage{makecell}
\usepackage{soul}
\usepackage{booktabs}

\newcolumntype{Y}{>{\centering\arraybackslash}X}

\begin{document}

\title{Dynamical formation of black hole binaries in dense star clusters: Rapid cluster evolution code}

\begin{abstract}
Gravitational-wave observations have just started probing the properties of black hole binary merger populations. 
The observation of binaries with very massive black holes and significantly asymmetric masses motivates the study of dense star clusters as astrophysical environments which can produce such events dynamically.
In this paper we present {\tt Rapster} (for ``Rapid cluster evolution''), a new code designed to rapidly model binary black hole population synthesis and the evolution of massive star clusters based on simple, yet realistic prescriptions. 
We also perform a thorough comparison with the Cluster Monte Carlo code and find generally good agreement.
The code can be used to generate large populations of dynamically formed binary black holes.
\end{abstract}

\author{Konstantinos Kritos}
\email{kkritos1@jhu.edu}
\affiliation{Department of Physics and Astronomy, Johns Hopkins University, 3400 N. Charles Street, Baltimore, Maryland, 21218, USA}

\author{Vladimir Strokov}
\email{vstrokov1@jhu.edu}
\affiliation{Department of Physics and Astronomy, Johns Hopkins University, 3400 N. Charles Street, Baltimore, Maryland, 21218, USA}

\author{Vishal Baibhav}
\email{vishal.baibhav@northwestern.edu}
\affiliation{Center for Interdisciplinary Exploration and Research in Astrophysics (CIERA) and Department of Physics and Astronomy, Northwestern University, 1800 Sherman Avenue, Evanston, Illinois 60201, USA}

\author{Emanuele Berti}
\email{berti@jhu.edu}
\affiliation{Department of Physics and Astronomy, Johns Hopkins University, 3400 N. Charles Street, Baltimore, Maryland, 21218, USA}

\date{\today}
\maketitle

\tableofcontents

\section{Introduction}
\label{sec:Introduction}

The detection of gravitational waves (GWs) in 2015~\cite{LIGOScientific:2016aoc} ushered in a major revolution in modern astrophysics and cosmology.
The GW transient catalogs published by the LIGO-Virgo-KAGRA collaboration can be used to carry out population studies and infer the properties of neutron stars (NSs) and black holes (BHs)~\cite{LIGOScientific:2021psn}.
The intrinsic characteristics predicted by astrophysical models can also be compared with those of the observed events to learn about the environments that host these mergers.

A key question concerns the origin of the observed binary black hole (BBH) mergers.
There are currently two main astrophysical scenarios proposed for the formation of BBHs, which involve either isolated binaries in the field or dynamical assembly in star clusters~\cite{Mandel:2018hfr,Mapelli:2021taw,Spera:2022byb}.
In turn, each scenarios could consist of multiple sub-channels, such as common envelope evolution~\cite{Dominik:2012kk,Broekgaarden:2021efa}, stable mass transfer~\cite{Olejak:2022zee}, and chemically homogeneous evolution~\cite{Riley:2020btf} for the isolated binaries in the field; and formation in young massive star clusters~\cite{DiCarlo:2019pmf,Banerjee:2017mgr,Banerjee:2019ute,Banerjee:2021wqh,Ziosi:2014sra,Rizzuto:2021atw,Sedda:2021abh}, globular clusters~\cite{Kulkarni:1993fr,PortegiesZwart:1999nm,Rodriguez:2016kxx,Samsing:2017xmd,Rodriguez:2019huv,Askar:2016jwt,Arca-Sedda:2018qgq}, nuclear star clusters~\cite{Miller:2008yw,OLeary:2008myb,Antonini:2016gqe,Hoang:2017fvh,Fragione:2018yrb,Mapelli:2021syv,Wang:2020jsx,Sedda:2020jvg}, open clusters~\cite{Kumamoto:2018gdg,Rastello:2021gvw,Rastello:2018elx}, and the disks of active galactic nuclei~\cite{Samsing:2022fxi,Tagawa:2019osr} for the dynamical channel.

At the time of writing, it is highly uncertain how much each of these channels actually contributes to the intrinsic astrophysical merger rates and to the overall population of BBH mergers observed in GWs~\cite{Belczynski:2021zaz,Sedda:2020vwo}.
Some studies claim that a single formation channel could still explain all of the observed events~\cite{Belczynski:2020bca,Rodriguez:2021nwd} because several poorly constrained parameters appear in the astrophysical modeling of each formation channel, and this leads to large uncertainties in the predicted properties of the single events and of the overall population.
The general consensus is that a combination of various formation scenarios, including possibly also primordial BHs~\cite{Bird:2016dcv}, is more likely~\cite{Stevenson:2015bqa,Vitale:2015tea,Stevenson:2017dlk,Zevin:2017evb,Bouffanais:2019nrw,Zevin:2020gbd,Wong:2020ise,Franciolini:2021tla}: see e.g.~\cite{Mandel:2021smh} for a recent review.
Given a sufficiently large number of events, it might be possible to use statistical methods (such as Bayesian inference) to disentangle isolated and dynamical mergers by comparing the intrinsic binary parameters measured in GWs (e.g., the component masses and spins) with the predictions of population synthesis models.

In this work, we model and simulate the assembly of BBHs in dynamical environments with a new open source Python code, \texttt{Rapster}\,\footnote{The code and documentation are available on \texttt{github} at the URL \url{https://github.com/Kkritos/Rapster}.}  (for ``Rapid cluster evolution'').

The development of this code is mainly motivated by the necessity to simulate a large number of clusters within a reasonable time.
There are several state-of-the-art numerical packages that can provide accurate results (including direct $N$-body solvers~\cite{2003gnbs.book.....A,Banerjee:2009hs,2016MNRAS.458.1450W,Sedda:2023qlx} or the {\tt CMC}~\cite{Rodriguez:2021qhl,Kremer:2019iul,Rodriguez:2019huv} and {\tt MOCCA}~\cite{Giersz:2011em,Maliszewski:2021jci} codes, which rely on a Monte Carlo approach based on H\'enon's method to evolve self-gravitating million-body systems). New implementations of BBH mergers and stellar evolution have recently been added to $N$-body codes~\cite{Banerjee:2019jjs,2022MNRAS.516.3266K,Rizzuto:2021atw,Rizzuto:2022fdp}. Despite these advances, it is computationally expensive to apply these techniques to produce large BBH populations or to infer astrophysical hyperparameters, due to the large number of star clusters that must be simulated.
In contrast, \texttt{Rapster} can simulate the evolution of star clusters within seconds for light clusters ($M_{\rm cl}<10^6M_\odot$), or within a minute for more massive clusters.
The simulation runtime grows for the heaviest nuclear star clusters---very dense, massive systems found at the centers of most galaxies and comprising $\sim 10^8$ bodies or more.

A second motivation is the detection of GW events with asymmetric mass-ratio  (e.g., GW190412~\cite{LIGOScientific:2020stg} and GW190814~\cite{LIGOScientific:2020zkf}) and of BBH mergers such as GW190521~\cite{LIGOScientific:2020iuh}, in which the binary components have masses within or above the so-called pair-instability supernova (upper) mass gap~\cite{Heger:2001cd,Woosley:2007qp,Belczynski:2016jno,Woosley:2016hmi}.
These massive BH components could themselves result from hierarchical mergers~\cite{Gerosa:2017kvu,Fishbach:2017dwv,Gerosa:2021mno,Fragione:2023kqv}: in particular, BHs could assemble dynamically in a cluster, merge and be retained, so that they can merge again in the same dense environment~\cite{Baibhav:2020xdf,Samsing:2020qqd}. To obtain the BH population in hierarchical merger scenarios and compare it with GW observations, it is necessary to simulate their dynamical assembly in multiple clusters across the Universe. This task is computationally prohibitive for most existing direct or indirect $N$-body codes.
It should be noted however that there are several proposed alternatives to repeated BH mergers that could produce BHs within the upper-mass gap, including runaway stellar collisions~\cite{DiCarlo:2019fcq,Kremer:2020wtp,Banerjee:2021wqh}, tidal-disruption events~\cite{Rizzuto:2022fdp}, or even primordial BHs~\cite{Kritos:2020wcl}.

\texttt{Rapster} is based on a semianalytic approach relying on a set of simple, yet realistic prescriptions (see Sec.~\ref{sec:Dynamics}).
Unlike existing rapid semianalytic codes to evolve BBHs in star clusters~\cite{Mapelli:2021gyv,Choksi:2018jnq,Antonini:2019ulv,Fragione:2021nhb} which use scaling relations, in our code we treat all dynamical channels that occur simultaneously in the cluster as Poisson processes.
Other semianalytic codes similar to {\tt Rapster} include {\tt FASTCLUSTER}~\cite{Mapelli:2021gyv}, {\tt cBHB}~\cite{Antonini:2019ulv}, {\tt B-POP}~\cite{Sedda:2021vjh}, and {\tt QLUSTER}~\cite{Gerosa:2023rwu}.
We compute the initial single-BH mass spectrum using  {\tt SEVN}~\cite{Spera:2017fyx} to find the remnant mass as a function of the zero-age main-sequence (ZAMS) mass of massive progenitor stars, but the code is modular, and any other initial mass spectrum (e.g. from {\tt SSE}~\cite{Hurley:2000pk,Banerjee:2019jjs}) can be used as input. To compute the properties of merger remnant products, such as mass, spin and GW kick, we use the {\tt precession} code~\cite{Gerosa:2016sys,Gerosa:2023xsx}.

The plan of the paper is as follows.
In Sec.~\ref{sec:Dynamics} we summarize the semianalytic prescriptions implemented in the code. 
In Sec.~\ref{sec:Comparison_with_CMC} we show comparisons of \texttt{Rapster} results with the Cluster Monte Carlo ({\tt CMC}) code.
In Sec.~\ref{sec:conclusions} we present our conclusions.

Throughout the code and in this text, we use astrophysical units in which we measure masses in $M_\odot$, distances and radii in pc (and occasionally in AU), velocities in $\rm km\ s^{-1}$, and time in Myr.
In those units the gravitational constant is approximately $G\simeq(232)^{-1}$~\cite{2003gmbp.book.....H}.

\section{Dynamical model}
\label{sec:Dynamics}

This section is dedicated to the physical model underlying {\tt Rapster}. We first summarize our treatment of star cluster evolution. Then we describe how we populate clusters with BHs, how we treat binary--single interactions, and the various dynamical channels that can lead to the formation of BBHs. We conclude with a flow chart that illustrates our simulation algorithm.

\subsection{Star cluster evolution}
\label{sec:Star_cluster_evolution}

Star clusters are formed from the fragmentation of giant molecular clouds. The majority of stars (if not all) form in groups and larger associations~\cite{Lada:2003ss,2019ARA&A..57..227K}.
Since all stars in a cluster form at roughly the same time (with potential slight time delays resulting in different populations), they have similar chemical composition (or metallicity $Z$) at formation.
Stars will however form with different masses, whose distribution is assumed to follow the Kroupa~\cite{Kroupa:2000iv} initial mass function (IMF)---a broken power law, here assumed to be universal.
The lower end of the initial stellar masses is assumed to be $0.08M_\odot$ by default.
We implement the mean power law indices from Ref.~\cite{Kroupa:2000iv}, and a default power law index of $-2.3$ for stars heavier than the Sun.
For simplicity we ignore all finite size effects for stars, and treat all members of the cluster as point particles.
Clusters that avoid infant mortality, surviving the first phase of gas expulsion, collapse and virialize.

The root mean square velocity of stars in the cluster is determined by the virial theorem, and given by (see Ref.~\cite{1987degc.book.....S}, page~12)
\begin{align}
{\langle v_\star^2\rangle}^{1/2}&=\sqrt{0.4GM_{\rm cl}
\over r_{\rm h}}\nonumber\\&\simeq13\,{\rm km\, s^{-1}}\, \left({M_{\rm cl}\over10^5M_\odot}\right)^{1\over2}\left({r_{\rm h}\over1\,\mbox{pc}}\right)^{-{1\over2}},
\label{Eq.StarVelocity}
\end{align}
where $M_{\rm cl}$ is the total mass of the cluster, $r_{\rm h}$ is its half-mass radius, and $G$ is the gravitational constant.
It turns out that the numerical coefficient of 0.4 in the equation above depends weakly on the density profile, and it can vary by a factor of at most 2.
The stationarity condition above is assumed at every moment in time and it is a good approximation, because the timescale it takes for a star to cross the size of the cluster (roughly the time to reach virial equilibrium) is much smaller than the timescale it takes for the stellar distribution function to change~\cite{1985IAUS..113..521A}, also known as the relaxation timescale.

The subsequent evolution of an isolated cluster following core collapse is driven by its internal dynamics and is dominated by two-body relaxation and stellar mass loss. 
A fraction of about $\xi_{\rm e}\simeq0.0074$~\cite{1985IAUS..113..521A} of all stars in an isolated cluster\footnote{This result is obtained by integrating the normalized Maxwell-Boltzmann distribution with one-dimensional velocity dispersion parameter $\sigma$ from $2\sqrt{3}\sigma$ (the escape velocity) to infinity.} have velocities in the tail of the Maxwellian distribution, and escape the cluster's gravitational potential.
The tail is then replenished through close encounters, and more stars escape.
In addition, since star clusters are not isolated systems, the host galaxy affects the evolution as well.
The presence of an external tidal field enhances the mass loss rate from the cluster, because it creates a finite tidal boundary through which stars can escape. The tidal (or Jacobi) radius is given by $r_{\rm J}=(GM_{\rm cl}R_G^2/(3V_G^2))^{1/3}$, where $R_G$ and $V_G$ are the galactocentric radius and circular velocity, respectively. According to numerical experiments performed in~\cite{Gieles:2008ew} the dimensionless mass loss rate $\xi_{\rm e}$ should be modified by a multiplicative factor of $\simeq\exp(10r_{\rm h}/r_{\rm J})$.
As the cluster loses mass, the escape velocity $v_{\rm esc}=2\langle v_\star^2 \rangle^{1/2}$ (see Ref.~\cite{1987degc.book.....S}, page 51) decreases and more stars are likely to be ejected, so $M_{\rm cl}$ decreases exponentially.

The timescale on which energy is distributed throughout a collisional $N$-body system is controlled by the half-mass relaxation timescale (see Ref.~\cite{1971ApJ...164..399S}, Eq.~(30); and Ref.~\cite{1987degc.book.....S}, page 40)
\begin{align}
\tau_{\rm rh}&=0.138{N^{1\over2}r_{\rm h}^{3\over2}\over\overline{m}^{1\over2}G^{1\over2}\ln\Lambda}{1\over\psi}\nonumber\\
&\simeq128\,\mbox{Myr}\,\left({M_{\rm cl}\over10^5{M}_\odot}\right)^{1\over2}\left({r_{\rm h}\over1\,\mbox{pc}}\right)^{3\over2}{0.6{M}_\odot\over\overline{m}}{8\over\ln\Lambda}{1\over\psi},
\label{Eq:RelaxationTime}
\end{align}
where we have used $M_{\rm cl}=N\overline{m}$, and we set $\Lambda\approx0.02N$ in the Coulomb logarithm.
Here, $\overline{m}$ is the average stellar mass in the cluster, which is around $0.6{M}_\odot$ under the assumption of the Kroupa IMF in the range $[0.08,150]M_\odot$.
Moreover, $\psi$ is the dimensionless mass moment factor, defined as $\overline{m^{5/2}}/\overline{m}^{5/2}$, and accounts for a cluster composed of multiple mass components.
It has been demonstrated numerically that multimass systems evolve faster due to a smaller relaxation timescale~\cite{1995ApJ...443..109L}. 
For two-mass models with a dominant component (such as one that contains stars and BHs) we can write $\psi=1+S$, where the symbol $S$ represents the Spitzer parameter, accounting for the effect of the BH subcluster~\cite{Antonini:2019ulv}.
The relaxation time is smaller than the lifetime of collisional systems; in particular, this criterion is met by star clusters. 
This is in contrast to stars in the solar neighborhood, where the relaxation time is much larger than the Hubble time: this makes galactic fields effectively collisionless, so that dynamical encounters play no role.
Lighter clusters have smaller relaxation times and evolve faster than massive systems~\cite{2010MNRAS.408L..16G}.

We describe the internal and external evolution of star cluster environments as in Refs.~\cite{Antonini:2019ulv,Gnedin:2013cda,Gieles:2008ew}. 
A code that simulates the global evolution of star clusters in a manner similar to ours is {\tt EMACSS}~\cite{2012MNRAS.422.3415A,2014MNRAS.442.1265A}.
If we write $M_{\rm cl}=N\overline{m}$, then the rate of change of the cluster's mass is due to variations in $\overline{m}$ (mass loss due to stellar evolution) and in $N$ (relaxational loss due to ejections).
Since it takes approximately one half-mass relaxation timescale for the system to thermalize and for the unbound high-mass tail of the Maxwellian to be filled, we model the cluster mass loss due to relaxation processes as 
\begin{align}
	{dM_{\rm cl}^{\rm (rlx)}\over dt}\equiv\overline{m}{dN\over dt} = - {\xi_{\rm e}M_{\rm cl}\over \tau_{\rm rh}}.
	\label{eq:rlx_mass_term}
\end{align}
Note that we have ignored the factor of 2.5 in $\xi_{\rm e}$ in the last equation above (which is accounted for in~\cite{Gnedin:2013cda}) because the effect of a multimass distribution has already been included through the mass moment $\psi$ in the expression for $\tau_{\rm rh}$ (see Eq.~\eqref{Eq:RelaxationTime}).
Finally, the cluster loses mass as a consequence of stellar evolution and winds according to~\cite{Antonini:2019ulv,2014MNRAS.442.1265A}:
\begin{align}
	{dM_{\rm cl}^{\rm (sev)}\over dt} \equiv N{d\overline{m}\over dt} = -\nu{M_{\rm cl}\over t}\Theta(t-\tau_{\rm sev}),
	\label{eq:sev_mass_term}
\end{align}
where $\nu=0.07$, $\tau_{\rm sev}=2\, \rm Myr$ and $\Theta(x)$ denotes the Heaviside function.
In deriving this previous equation, we assume the average mass to evolve according to $\overline{m}=\overline{m}_0(t/\tau_{\rm sev})^{-\nu}$ for $t>\tau_{\rm sev}$~\cite{2014MNRAS.442.1265A}, where $\overline{m}_0$ is the initial average mass.
The total rate of change of $M_{\rm cl}$ is then determined by adding the relaxation and stellar evolution terms in Eqs.~\eqref{eq:rlx_mass_term} and~\eqref{eq:sev_mass_term}, respectively.
In $dM_{\rm BH}/dt$ we do not include the BH mass loss rate, because the latter contributes a very small amount in systems that are dominated by low-mass stars, as is the case in systems that follow the default Kroupa IMF.
In our simulations, the BH ejection rate is determined by the microphysical processes (primarily binary--single interactions and relativistic recoils) that occur in the core of the system and will be discussed in a later section (Sec.~\ref{sec:BH_ejections}).

Energy production in the core of the cluster, originating from the formation of tight binaries and from the interactions of stars with tight binaries, causes the cluster to slowly expand.
We model the time evolution of the half-mass radius due to relaxation and stellar evolution processes by writing~[see~\cite{Antonini:2019ulv}, Eq.~(8), (15), and~(17)]:
\begin{align}
	{dr_{\rm h}\over dt} = \left[\zeta{r_{\rm h}\over\tau_{\rm rh}} + 2{dM_{\rm cl}^{\rm (rlx)}\over dt}{r_{\rm h}\over M_{\rm cl}}\right]\Theta(t - \tau_{\rm cc}) - {dM_{\rm cl}^{\rm (sev)}\over dt}{r_{\rm h}\over M_{\rm cl}}.
	\label{eq:half_mass_radius_evolution_equation}
\end{align}
The first term in this equation becomes effective for times $t>\tau_{\rm cc}$, where $\tau_{\rm cc}=3.21\tau_{\rm rh,0}$ is the core collapse timescale~\cite{Antonini:2019ulv} and $\tau_{\rm rh,0}$ is the initial half-mass relaxation time, because it is only after the core collapses that energy is generated via binary formation and interactions.
We set the dimensionless constant to an average value of $\zeta=0.08$ (for both the whole cluster and the BH subsystem) to match numerical simulations of tidally limited clusters~\cite{2012MNRAS.422.3415A,Breen:2013vla,2011MNRAS.413.2509G}.
Here, $\zeta$ represents the fraction of the total energy of the cluster that can be conducted by way of two-body relaxation through $r_{\rm h}$ and shared among the members of the cluster within one $\tau_{\rm rh}$.
In other words, the heat flow rate via gravitational encounters in a star cluster is $\simeq0.2\zeta GM_{\rm cl}^2/(r_{\rm h}\tau_{\rm rh})$.

Finally, we implement Eqs.~(7) and~(8) from~\cite{Gnedin:2013cda} to evolve the galactocentric radius of the cluster under the effect of dynamical friction.
As the cluster inspirals toward the galaxy's center because of dynamical friction, it can be tidally stripped or merge with the central nuclear star cluster (if present).

In implementing all the above mentioned differential equations (along with Eq.~\eqref{eq:half_mass_radius_evolution_equation},~\eqref{eq:sev_mass_term}, and~\eqref{eq:rlx_mass_term} above) we use a finite difference scheme. We update the half-mass radius and cluster mass by choosing the increment $dt$ to match the time step of the simulation. (See Sec.~\ref{sec:GlobalSimulationDetails} for a discussion of the adaptive time step used in our simulations.)

\subsection{Black holes in star clusters}
\label{sec:BHinClusters}

To populate clusters with BHs, we need to introduce prescriptions for their initial mass (which is related to the ZAMS mass of the progenitor stars), their spin, the kick they receive at birth, and their distribution in the cluster.

\subsubsection{Remnant-mass prescription}
\label{sec:Remnant-mass_prescription}

Massive stars evolve quickly and produce compact object remnants within a few million years.
It is expected that a population of BHs might reside near the centers of star clusters. This hypothesis is supported by spectroscopic and kinematic observational evidence for the existence of compact objects in dense star clusters~\cite{Maccarone:2007dd,Strader:2012wj,2018MNRAS.475L..15G,2019A&A...632A...3G,Vitral:2022apu}.
We implement the compact remnant mass--ZAMS  mass prescription based on the Fryer et al. (2012) delayed model~\cite{Fryer:2011cx} as a default option.
Other remnant mass models included are the Fryer {\it et al.} (2012) rapid function based on the {\tt SSE} code~\cite{2000MNRAS.315..543H}, and the delayed model of the {\tt SEVN} code~\cite{Spera:2017fyx}.
The implementation is not tied to a specific stellar evolution code, allowing easy integration of alternative prescriptions for remnant mass based on ZAMS mass and metallicity. To enhance flexibility, the code offers customization through an input option for users to provide their own list of BH masses retained in the cluster.
For the Fryer {\it et al.} (2012) rapid and delayed models we use the {\tt updated-BSE} code described in Ref.~\cite{Banerjee:2019jjs}. For computational efficiency, we create look-up tables with remnant masses on a $700\times 700$ grid in ZAMS mass vs. metallicity, and we perform a two-dimensional interpolation of this grid.

The {\tt SEVN} code allows us to obtain BH masses as a function of the ZAMS mass and metallicity of the progenitor stars, $m_{\rm rem}=m_{\rm rem}(m_{\rm ZAMS},Z)$. 
The implementation of this procedure in {\tt SEVN} works by interpolating stellar evolutionary tracks from {\tt PARSEC} simulations~\cite{2014MNRAS.445.4287T}, and it includes the latest stellar wind and pair-instability supernova prescriptions.
For the supernovae, we adopt the delayed core-collapse engine, although the exact details of the supernova convection efficiency affect only the low-mass end of the BH mass spectrum~\cite{Fryer:2011cx,Olejak:2022zee}.
This choice gives rise to a number of BHs within the so-called ``lower mass gap'' that ranges from $\simeq2.5M_\odot$ to $\simeq5M_\odot$~\cite{2011ApJ...741..103F}.
We calculate the remnant masses only for stars with ZAMS masses above $20M_\odot$, because in {\tt SEVN} these stars produce remnants above $\sim3M_\odot$ regardless of metallicity: see e.g. Fig.~2 of Ref.~\cite{Spera:2017fyx}.
Then, we keep track of only those BHs that are more massive than 3$M_\odot$.
This gives us the total number of BHs originally produced in the cluster, $N_{\rm BH,0}^{\rm tot}$.
The interpolation provided by {\tt SEVN} handles metallicities in the (absolute) range between~$10^{-4}$ and~$0.02$.

In environments with metallicity lower than $\approx5\%$ of the solar metallicity, which we assume to be $Z_\odot=1.4\%$~(\cite{2009ARA&A..47..481A}, page~24), very massive stars presumably collapse directly, avoiding the pair-instability mechanism and producing BH remnants with masses above the upper mass gap.
We can account for these direct-collapse BHs by extrapolating the IMF up to stars with a ZAMS mass of 340$M_\odot$, which is the heaviest star the {\tt SEVN} code interpolates based on
simulations~\cite{2015MNRAS.452.1068C}.
This provides one of the possible formation pathways for intermediate-mass BHs (IMBHs): the direct collapse of low-metallicity stars at high redshift can generate a few BHs with masses above 100$M_\odot$ in star clusters, which can then further grow by mergers~\cite{Spera:2017fyx,Mangiagli:2019sxg} or accretion (see e.g.~\cite{Natarajan:2020avl}).
Despite the typically chosen maximum ZAMS mass of $150M_\odot$ employed in theoretical studies (a value based on observations of stars in the Arches cluster~\cite{Figer:2005gr}), cosmological simulations provide numerical support  to the idea that zero-metallicity stars could reach hundreds of solar masses~\cite{Susa:2014moa}.
Moreover, spectroscopic observations and $N$-body simulations for the star cluster R136 in the Large Magellanic Cloud suggest the existence of stars more massive than the putative theoretical limit of $150M_\odot$~\cite{2010MNRAS.408..731C} (but see~\cite{2012ApJ...746...15B} for interpretations of those stars as the result of stellar runaway mergers).
Motivated by these considerations, and taking into account the uncertainty in the IMF for very massive stars, we set the maximum ZAMS mass to be a free parameter that can be provided as an input to our code. We choose the highest possible value for this parameter to be~$340M_\odot$, due to the lack of simulations for more massive stars.
We have checked that varying this parameter from $150M_\odot$ to $340M_\odot$ changes the number of BH progenitors at the percent level. The initial number of BHs is expected to vary even less at low metallicities, because most stars above $150M_\odot$ completely explode through pair-instability supernova (PISN), leaving behind no BH remnant.

\subsubsection{BH spins}

We consider two prescriptions for the dimensionless spin parameter $\chi_{\rm 1g}\in[0,1]$ of ``first-generation'' BHs (i.e., those formed by stellar collapse).
Some stellar models favor low natal spins for first generation BHs~\cite{Fuller:2019sxi}, but we opt to keep this parameter as an input, especially because individual spins are difficult to measure and strongly constrain with current GW observatories~\cite{LIGOScientific:2021djp,Callister:2022qwb}. One option is a monochromatic (delta function) distribution at a given $\chi_{\rm 1g}$; we also consider another simple option, i.e., a uniform distribution in the range $[0,\chi_{\rm 1g}]$. It is possible to modify the source code to include different distributions for the spin of 1g BHs.

Whenever the spins $\boldsymbol{\chi}$ of BHs in a binary are nonzero, we randomize their orientations by sampling $\boldsymbol{\chi}\cdot\boldsymbol{L}/|\boldsymbol{L}|\equiv\cos\theta$ uniformly in $[-1,1]$. Here, $\boldsymbol{L}$ stands for the orbital angular momentum of the BBH, and $|\boldsymbol{L}|$ denotes its magnitude.
The relative angle $\Delta\phi$ between the spin projections in the orbital plane is sampled uniformly in $[0,2\pi]$.
Since the spins can undergo precession as the BBH goes through the inspiral, we assume the angles to be defined at the last stable circular orbit, just before the final plunge and coalescence of the binary. However, since an initially isotropic distribution evolves to an isotropic distribution~\cite{Schnittman:2004vq,Kesden:2014sla,Gerosa:2015tea}, the frequency at which these angles are defined does not really matter.
It is possible for the user to alter these choices by modifying the function {\tt sample$\_$angles()} in the {\tt functions2.py} file for a nonisotropic sampling of the spin directions (such as clusters with a disklike geometry, that tend to have a preferred orientation in space~\cite{Santini:2023ukl}).

After the merger of two BHs, we compute the spin of the merger product (as well as other parameters, such as the GW-induced recoil and remnant mass) with fitting formulas implemented in the {\tt precession} code~\cite{Gerosa:2016sys,Gerosa:2023xsx}.
Those formulas are fitted to numerical relativity simulations of merging BBHs and based on Refs.~\cite{Campanelli:2007ew,Gonzalez:2006md,Lousto:2007db,Lousto:2012su,Lousto:2012gt,Gonzalez:2007hi} for the GW recoil, Ref.~\cite{Barausse:2012qz} for the final mass, and Ref.~\cite{Barausse:2009uz} for the final spin.

\subsubsection{Natal kicks}

It is believed that BHs and neutron stars generally have nonzero velocity (receive a ``kick'') at birth as a consequence of asymmetric supernova mass ejection and the conservation of momentum~\cite{1994Natur.369..127L,Janka:2013hfa}.
If that kick exceeds the escape velocity of the host environment, the remnant escapes the gravitational pull of the cluster, and no longer contributes to internal dynamical processes.
This means that only a fraction $f_{\rm ret}$ of the BHs is retained in the cluster.
The recoil has been constrained to be a few hundreds of $\rm km\, s^{-1}$ for neutron stars, based on observations of pulsar proper motions in the Milky Way~\cite{Hobbs:2005yx,Kapil:2022blf}. For BHs, the recoil velocity is highly uncertain~\cite{Mandel:2015eta,Belczynski:2015tba}.
It is believed, however, that massive stellar cores that directly collapse into heavy BHs receive little or no kick at birth, due to material falling back onto the newly born compact remnant.

We calculate the fallback (``fb'') fraction $f_{\rm fb}$ of the ejected supernova mass that falls back onto the BH by implementing  Eq.~(19) and~(16) from Ref.~\cite{Fryer:2011cx} in the case of the delayed and rapid core-collapse engine, respectively.
Whenever required, we use the analytic formulas from Ref.~\cite{Hurley:2000pk} to determine the carbon/oxygen (CO) core mass when we use the Fryer et al. (2012) models. When instead we use {\tt SEVN}, we consistently use the CO core mass output of the {\tt SEVN} code (see~\cite{Spera:2017fyx}).
To speed up the simulations, we have also precomputed the CO core mass output of {\tt SEVN} using the same grid of ZAMS mass and metallicity as for the remnant masses (see Sec.~\ref{sec:Remnant-mass_prescription}). This output is saved in look-up tables whose values are later interpolated.
The BH kick at birth, $v_{\rm kick,0}$, is originally drawn from a Maxwellian distribution with one-dimensional root-mean-square velocity of $265\,\rm km\, s^{-1}$~\cite{Hobbs:2005yx} (in our code, the user can set this parameter to a value different from the default).
Due to isotropy, the 3-dimensional natal kick is obtained by multiplying the sampled velocity by $\sqrt{3}$.
The final natal kick is then determined by either $v_{\rm kick}=(1-f_{\rm fb})v_{\rm kick,0}$ in the fallback prescription, or (by default) $v_{\rm kick}=(1.4M_\odot/m_{\rm BH})v_{\rm kick,0}$ in the momentum conservation prescription, where $m_{\rm BH}$ is the mass of the BH~\cite{Giacobbo:2019fmo}.
Note that $f_{\rm fb}$ depends on the ZAMS mass $m_{\rm ZAMS}$ of the progenitor star: it first increases linearly with $m_{\rm ZAMS}$ (see Eq.~(2) from Ref.~\cite{Belczynski:2005mr}) until it takes the value $f_{\rm fb}=1$ when $m_{\rm ZAMS}>35$M$_\odot$.
Therefore ``heavy'' BHs with masses above $\sim9M_\odot$ (the exact value of this critical mass depending on metallicity) receive small or no kick at birth~\cite{Belczynski:2007xg}, and they tend to be retained even in light clusters with lower escape velocities.

\subsubsection{BH segregation, velocities, and core radius}
\label{sec:SegragationAndVelocities}

As a result of interactions that occur in the dense cluster, energy is distributed among the members of the system, with BHs slowing down and lighter objects gaining kinetic energy via two-body encounters.
As relics of massive star evolution, BHs become the heaviest components in the cluster, and if not already formed in the core, they sink into the central regions via dynamical friction.
This process, known as mass segregation, occurs on a relatively short characteristic timescale, which is a fraction $\simeq\overline{m}/m_{\rm BH}$ of the relaxation time for a heavy tracer of mass $m_{\rm BH}$~\cite{Fregeau:2001rp}.
As a result, the BHs condense in a subcluster with half-mass radius $r_{\rm h, BH}$ smaller than the half-mass radius of the whole system, $r_{\rm h}$.

As BHs cool down and stars heat up, the former sink deep into the core, and the temperature ratio $\xi\equiv m_{\rm BH}\langle v_{\rm BH}^2\rangle/(\overline{m}\langle v_\star^2\rangle)$ approaches unity.
The system thus evolves toward a state of energy equipartition, which is, however, not always achieved in realistic $N$-body systems with a continuous mass spectrum, as shown both analytically and numerically~\cite{1969ApJ...158L.139S,1978ApJ...223..986V,Khalisi:2006ne,2013MNRAS.435.3272T}.
The impossibility of thermal and dynamical equilibrium is expressed by the fact that the temperature ratio asymptotically stabilizes at some minimum value $\xi_{\rm min}>1$ which depends solely on the individual ($p\equiv m_{\rm BH}/\overline{m}$) and total ($P\equiv M_{\rm BH}/M_{\rm cl}$) mass ratios in two-mass models~\cite{2000ApJ...539..331W,Breen:2013vla}.
Here, $M_{\rm BH}$ stands for the total mass of the BH subsystem in the cluster.
An approximate expression for this minimum temperature ratio between BHs and stars in the cluster is $\xi_{\rm min}=p^{3/5}P^{2/5}(\ln\Lambda_{\rm BH}/\ln\Lambda)^{-2/5}$, where $\Lambda_{\rm BH}=0.02N_{\rm BH}$ is the Coulomb logarithm of the BH subsystem. 
To avoid numerical divergences when the BH subsystem evaporates, we set $\Lambda_{\rm BH}=1$ and $\Lambda=10$, as an approximation.
This expression for $\xi_{\rm min}$ is derived in the framework of the Breen and Heggie theory by assuming a balanced evolution of the cluster and equating the energy fluxes through the BH and the stellar half-mass radii~\cite{Breen:2013vla}.
Moreover, we write the Spitzer parameter as $S=p^{3/2}P$, and as $S=pP$ when the system is Spitzer stable (i.e., when $p^{3/2}P<0.16$)~\cite{Antonini:2019ulv}.
If evaluating $\xi_{\rm min}$ returns a result that is less than unity, we set $\xi_{\rm min}=1$, because in that case equipartition can be achieved.

Since the equipartition is not always attainable, BHs continue to collapse to form a densely packed, self-gravitating, and partially decoupled subsystem, which dominates the dynamics near the center.
Applying the virial theorem, the 3-dimensional velocity dispersion in the BH subsystem is $\langle v_{\rm BH}^2\rangle=0.4GM_{\rm BH}/r_{\rm h,BH}$, where $r_{\rm h,BH}$ is the segregation radius of BHs.
Combining this equation with the definition of the temperature ratio and Eq.~\eqref{Eq.StarVelocity}, we obtain an expression for the segregation radius of BHs:
\begin{align}
{ r_{\rm h,BH} \over r_{\rm h} } \simeq 0.04\times {4\over\xi_{\rm min}} \left( {m_{\rm BH}\over 10{M}_\odot} \right)^2 {N_{\rm BH}\over1000} {0.6{M}_\odot\over \overline{m}} {10^6{M}_\odot \over M_{\rm cl}} .
\label{Eq.segRadius}
\end{align}
This quantity defines the size of the compact region where most heavy BHs reside~\cite{Lee:1994nq}.
The parameterizations of segregation radius and velocity dispersion, Eq.~\eqref{Eq.StarVelocity}, with their dependence on mass play an important role, because they significantly affect collision rates and the BBH dynamical assembly timescale.

The energy flux through $r_{\rm h,BH}$ is balanced by the energy generation rate in the core of the BH subcluster~\cite{Breen:2013vla}. 
If those two rates are not equal, then the core will adjust accordingly by either expanding or contracting to meet the energy demands of the whole system. 
Energy can flow in the cluster at a rate set by the efficiency of two-body relaxation (see the discussion in Sec.~\ref{sec:Star_cluster_evolution}). 
In particular, if there are no hard binaries in the cluster (the main source of energy in the systems of interest in this work) then the core will contract and the central BH density will increase until hard binaries form, and heat production through binary--single interactions reverses the collapse.
During this phase the core collapses isothermally, because the bulk of the BH subsystem acts as a heat bath that keeps the one-dimensional velocity dispersion $\sigma_{\rm BH}$ roughly fixed.
During balanced evolution, the core radius of the BH subsystem is given by~\cite{Breen:2013vla}
\begin{align}
	{r_{\rm c, BH} \over r_{\rm h,BH}} = \left({{\cal C}\over\zeta N_{\rm BH}^2\ln\Lambda_{\rm BH}}\right)^{1/3}.
	\label{Eq.BHcoreRadius}
\end{align}
The variable ${\cal C}$ is a constant that depends on the macro- and microproperties of the cluster (see Appendix~\ref{app:BHcoreRadius} for details).
The theory breaks down when the number of BHs decreases below some threshold which is around a few tens of BHs, at which point the core and half-mass radii of the BH subsystem become comparable.

The formation process of young massive clusters observed in the local Universe does not exceed $1$~Myr. This conclusion is based on the study of the young heavy stellar populations that those systems harbor, and the fact that they are observed to be virialized~\cite{2014CQGra..31x4006K}.
In addition, before the BHs segregate to form the BH subsystem and the core collapses, no significant dynamical interactions take place due to the reduced BH densities, as evidenced by $N$-body simulations~(see Fig.~2 of~\cite{2010MNRAS.402..371B}).
Moreover, we form all BHs in the system at $\simeq3.5~\rm Myr$ after cluster formation, which is on average the lifetime of stars more massive than $20M_\odot$ (see Table~1 of~\cite{Woosley:2002zz}).
In reality, heavier stars collapse first and BH formation spans a time interval of a few Myr, which we neglect.
Therefore, we simplify our model by starting off all dynamical processes with an initial condition consisting of a cluster in virial equilibrium, with the retained BH remnants segregated into its core after their formation.
We then evolve such a system in time according to the prescriptions presented above.
The initial phases of cluster evolution (spanning approximately a few million years) are ignored as they are hard to model and involve an interplay between gas physics, stellar evolution, and feedback~\cite{2010ARA&A..48..431P}.
Furthermore, we take the central density of stars $\rho_{\rm c}$ to be uniform in the core of the cluster, an assumption in agreement with the luminosity curves of star clusters~\cite{1962AJ.....67..471K}. 
For computational convenience, we also consider a uniform spatial distribution for the BHs within their core radius.
As the cluster expands, the central density is self-similarly evolved according to $\rho_{\rm c}=\rho_{\rm c,0}(M_{\rm cl}/M_{\rm cl,0})(r_{\rm h,0}/r_{\rm h})^3$, where a subscript~$0$ on any quantity denotes its initial value.
    
To reiterate, we treat all dynamical channels that occur simultaneously in a cluster as Poisson processes, each with its own timescale. In what follows we provide the reaction rates for each relevant physical process. 

\subsection{Binary--single interactions}
\label{sec:CollisionTimeAndHardening}

In this section we describe all semianalytic prescriptions we use when simulating the interaction of a BBH with a third body.
Let us consider a binary with mass components $m_1$, $m_2$ and semimajor axis $a$.
We define the ``hardness ratio'' $\eta$ to be the ratio of the binary's binding energy $Gm_1m_2/(2a)$ over the ambient kinetic energy of a typical BH in the cluster environment, $\overline{m}_{\rm BH}\langle v_{\rm BH}^2 \rangle/2$:
\begin{align}
\eta\equiv {Gm_1m_2\over\overline{m}_{\rm BH}\langle v_{\rm BH}^2\rangle a}.
\label{Eq:hardenessRatio}
\end{align}
We will call the binary ``hard'' if the hardness ratio exceeds unity, and ``soft'' otherwise.
By default we do not keep track of soft binaries with $\eta<1$ because these tend to be ionized according to Heggie's law.
In very massive clusters, where the velocity dispersion exceeds a few hundreds $\rm km\ s^{-1}$, there are no hard binary stars (not even contact binaries with the components on the main sequence), because the value of the hardness semimajor axis drops below the solar radius.

\subsubsection{Encounter timescale}

The average timescale for a specific target object to interact with a test particle with a maximum pericenter distance of interaction $r_p$ can be determined as the inverse of the collision rate: see Eqs.~(2.2)--(2.5) in Ref.~\cite{1993ApJ...415..631S}.
In the gravitational focusing regime, which is applicable as long as we are dealing with hard binaries, this timescale becomes
\begin{align}
\tau_{\rm enc}&\simeq1.6\,{\rm Myr}\nonumber\\&\times\left({m_{\rm tot}\over 20M_\odot}\right)\left({n\over10^5\,\mbox{pc}^{-3}}\right)\left({\langle v_{\infty}^2\rangle^{1/2}\over10\,{\rm km\, s^{-1}}}\right)^{-1}\left({r_{ p}\over10\,\mbox{AU}}\right),
\label{Eq:CollTime}
\end{align}
where $\sqrt{\langle{v}_\infty^2\rangle}$ is the relative root mean square velocity between the two bodies at infinity (not be confused with the one-dimensional velocity dispersion, which is a factor of $\sqrt{3}$ smaller); $m_{\rm tot}$ is the total mass of the two-body system; and $n$ is the number density of test particles.
Note that by ``target object'' or ``test particle'' we mean any type of multiplicity in the system, i.e., a single BH, single star, binary star, BH--star, BBH, or a BH triple.
For example, to calculate the average timescale it takes for a BBH to interact with another single BH we use the above formula, replacing $n$ by the number density of single BHs in their segregation volume.
When calculating the timescale of a binary--single interaction, the semimajor axis $a$ of the binary also sets the scale for the pericenter distance, in which case we set $r_p=2a$.
If we want to estimate the average timescale for an encounter to occur anywhere in the cluster between any two objects, we use Eq.~\eqref{Eq:CollTime} divided by the total number of target objects.
Most binary--single interactions among BHs occur in the core of the BH subsystem where the density is the highest, thus when applying Eq.~\eqref{Eq:CollTime} we replace $n$ by the core number density of BHs, $n_{\rm c,BH}\approx3N_{\rm BH}/(8\pi r_{\rm h,BH}r_{\rm c,BH}^2)$ under the assumption of an isothermal profile.

\subsubsection{Binary hardening}
\label{sec:binaryHardening}

Numerical and theoretical considerations show that hard binaries tend to increase their binding energy as a result of binary--single interactions, whereas soft binaries become softer with time as a consequence of the negative heat capacity of gravitational systems~\cite{Gurevich:1950,1975MNRAS.173..729H,1980AJ.....85.1281H}.
During interactions, on average, a hard binary gains an amount of binding energy $\Delta E_b$ that scales with the binding energy $E_b$ and with the mass ratio between the binary $m_{12}$ and the single $m_3$ as follows~\cite{1980AJ.....85.1281H,Antonini:2016gqe,Sesana:2006xw,Quinlan:1996vp}:
\begin{align}
	|\langle{\Delta E_b}\rangle|=\beta {m_3\over m_{12}}E_b.
	\label{Eq:Binding_energy_change}
\end{align}
The validity of the mass dependence of this formula has not yet been fully explored numerically, however we use it as an extrapolation between the test-particle and equal-mass cases.
After a binary--single interaction, the binary's Keplerian parameters (semimajor axis $a$ and eccentricity $e$) change to new values $a'$ and $e'$ according to
	\begin{align}
		{a'\over a}&=\left(1+\beta{m_3\over m_{12}} \right)^{-1},
	\end{align}
	\label{Eq.HardeningEq}%
while $e'$ is resampled from the thermal distribution.
Previous theoretical considerations~\cite{1937AZh....14..207A} and numerical experiments~(see~\cite{1975MNRAS.173..729H}, Sec.~2.3) have verified the validity of the thermal distribution for systems in thermal equilibrium.
The symbol $\beta$ in Eq.~\eqref{Eq:Binding_energy_change} above denotes a constant which we take to be $4/7$~\cite{Samsing:2017xmd}.

Depending on the pericenter of the encounter, the binary can be weakly or strongly perturbed. 
Since we take $r_p\le2a$, we only account for strong interactions. If the third body approaches and interacts closely with the members of the binary, the energy exchanged between the single and the binary will be comparable to the binding energy of the binary and a resonant interaction will occur.
During a resonant interaction multiple intermediate binary and three-body states will form before the system finally splits into a stable binary and a single~\cite{Samsing:2013kua}.
To determine whether the interaction is resonant we check whether the pericenter of the interaction $r_p$ is smaller or larger than $\max(m_1,m_2)a/m_{12}$~\cite{Samsing:2013kua}, which corresponds to the typical distance of the secondary from the BBH's center of mass.
Since the cross section $\pi b^2$ represents an area and the probability is proportional to the area, we sample the impact parameter uniformly in $b^2$.
In the gravitational focusing regime we have $\pi b^2\propto r_p$, hence we draw $r_p$ uniformly in the range $(0, 2a)$.

\subsubsection{Interaction recoil}
\label{sec:interactionRecoil}

A BBH interacting with a single object receives a recoil velocity.
From energy conservation considerations, one can relate the magnitude of the recoil velocity to the semimajor axis of the binary~\cite{1993ApJ...415..631S}.
Since the amount of energy extracted by the single is proportional to the binary's binding energy, harder binaries have higher recoil velocities.
Above some critical value of the hardness ratio, the value of the recoil velocity exceeds the escape velocity $v_{\rm esc}$ of the environment, and the BBH is ejected into the low-density field.
If we denote the reduced mass of the system by $\mu=(m_1+m_2)m_3/(m_1+m_2+m_3)$, the critical semimajor axis for BBH ejection after a binary--single collision is~\cite{Antonini:2016gqe}
\begin{align}
a_{\rm ej}&={\beta}{m_1m_2m_3\over (m_1+m_2)^3}{G\mu\over v_{\rm esc}^2}\nonumber\\&\simeq0.2\,\mbox{AU}\times{\beta\over0.8}{m_{\rm BH}\over10M_\odot}\left({v_{\rm esc}\over50\,{\rm km\, s^{-1}}}\right)^{-2},
\label{Eq.CritSMA4Ej}
\end{align}
where in the last approximation we assumed that all objects that participate in the binary--single interaction have the same mass: $m_1=m_2=m_3=m_{\rm BH}=10M_\odot$.
We have checked that our estimate of $a_{\rm ej}$ is consistent with Refs.~\cite{Rodriguez:2016kxx,Moody:2008ht}.
Had we chosen $m_1=m_2=10{M}_\odot$ and $m_3=0.6M_\odot$ (which is the mass of a typical star in the cluster) while keeping all other parameters the same, the critical value in Eq.~\eqref{Eq.CritSMA4Ej} would be $\simeq 0.001\,\rm AU$.
It is clear that BBHs have the highest probability of getting kicked out of the cluster before merger when they interact with heavier objects (here other single BHs) as a consequence of energy and momentum conservation.
This is because more massive bodies extract larger amounts of energy from the binary, and also because the final relative velocity is shared more evenly between the binary and the single.

If the BBH with mass components $m_1$ and $m_2$ interacts with another single BH of mass $m_3$, we then draw the mass of the third BH based on the dependence of the BBH--BH interaction rate on $m_3$.
In the gravitational focusing regime, the rate dependence on masses is $\tau_{\rm BBH-BH}^{-1} \propto(m_1+m_2+m_3)/\sigma_{\rm BBH,BH}$, where $\sigma_{\rm BBH,BH}=\sqrt{\sigma_{\rm BBH}^2 + \sigma_{\rm BH}^2}$ is the relative velocity dispersion between the BBH and the BH at infinity.
Noting that $\sigma^2\propto \xi/m$ and that $\xi\propto m^{3/5}$~\cite{Breen:2013vla} we find that
\begin{align}
	p(m_3)\propto{m_1+m_2+m_3\over\sqrt{(m_1+m_2)^{-2/5} + m_3^{-2/5}}},
\end{align}
is the probability density function for $m_3$, which we implement when sampling the mass of the single BH during a binary--single interaction.

As the BBH hardens, at some point it becomes so tight that the GW emission timescale becomes comparable to the encounter timescale, and the BBH may coalesce and produce a merger remnant within the cluster before getting ejected. If its Keplerian parameters happen to be such that the interaction timescale with another object in the cluster exceeds the GW coalescence timescale~\cite{Miller:2002pg}, the BBH enters into the gravitational radiation phase, where its orbital evolution is dominated by the release of GWs.
However, if the critical semimajor axis for ejection is reached before the BBH enters the GW regime (for instance, due to its interaction with another single BH), it escapes the cluster~\cite{Lee:1994nq}.
If that happens, we check whether the ejected BBH merges in isolation within the available time.
To compute the coalescence timescale of an eccentric BBH, we use the accurate analytical fit of Ref.~\cite{Mandel:2021fra}.

\subsubsection{BBH--BH exchanges}

When a star encounters a BBH, it typically extracts some amount of energy from the binary in a flyby.  The interaction of a BBH with a third BH is more interesting, especially if the latter is massive enough.  According to scattering experiments performed in~\cite{1980AJ.....85.1281H} (see e.g. their Fig.~4), the probability for the intruder to substitute one of the binary members is almost unity if its mass exceeds the component BBH masses by a factor $\simeq 1.6$.  
Furthermore the pericenter $r_p$ of the interaction must be comparable with the size of the binary $a$, otherwise the energy exchanged will not be enough to break the binary and facilitate the substitution.
Here, we simply perform a prompt exchange if $m_3>\min(m_1,m_2)$ and $r_p\le\max(m_1,m_2)m_{12}^{-1}a$, which is the condition for resonant interaction. In this case we trade the intruder for the lighter binary component, otherwise we ignore the substitution and regard the BBH--BH interaction as a hardening flyby episode. 
A more sophisticated treatment of BBH--BH exchanges would not significantly affect the final mass ratio distribution of BBH mergers.
The new BBH formed in the post-exchange state approximately conserves its binding energy, thus increasing its semimajor axis by a factor of $m_3/\min(m_1,m_2)$ relative to its previous value~\cite{1996ApJ...467..359H,1993ApJ...415..631S}, while its eccentricity is resampled from the thermal distribution. In other words, we assume that the binding energy of the new pair is the same as that of the old binary: if the old binary is hard, then so is the new one.
Once we perform the exchange interaction we also increase the binding energy of the new BBH according to Eq.~\eqref{Eq:Binding_energy_change}, because interactions involving hard binaries and singles are exothermic.

\subsubsection{BH ejections}
\label{sec:BH_ejections}

In our implementation, BHs are ejected as a consequence of the Newtonian recoil imparted into the binary and the single during strong BBH-BH interactions.
In such an encounter, we compute the recoil velocity of the BH and the BBH at the end of the interaction from the conservation of energy and linear momentum in one dimension. We decide whether to eject the BH and/or the BBH by comparing their recoil velocities with the escape velocity. 
If $m_3$ ($m_3'$) is the mass of the single and $m_{12}$ ($m_{12}'$) the mass of the binary before (after) the interaction, then the recoil velocity of the single, $v_{3}'$, and of the binary, $v_{12}'$, after the encounter are computed according to:
\begin{align}
v_3'{m_{123}\over m_{12}'}=v_{12}'{m_{123}\over m_3'}=\sqrt{\mu\over\mu'}\sqrt{v_{\rm rel}^2 + {2\over\mu}\Delta E_b},
\label{eq:recoil_velocities}
\end{align}
where $\Delta E_b$ is given by Eq.~\eqref{Eq:Binding_energy_change} and $v_{\rm rel}$ is the relative velocity at infinity before the interaction, $m_{123}=m_{12}+m_3=m_{12}'+m_3'$, and  $\mu$ ($\mu'$) is the reduced mass of the binary--single system before (after) the interaction.
Equations~\eqref{eq:recoil_velocities} are a consequence of energy and momentum conservation, assuming that the binary velocity $\boldsymbol{v}_{12}'$ and the single's velocity $\boldsymbol{v}_{3}'$ point in opposite directions.
Our formalism above is general and accounts for exchanges between a member of the binary and the incoming third BH; i.e., the interaction is of the form $(1-2)+3\to(1'-2')+3'$ in which the BHs in the final state may have been permuted.
In fact, since in our model $v_{3}'>v_{12}'$, the binary typically ejects the BH before ejecting itself from the cluster in cases where $m_3 < m_{12}$.

The energy released in a binary--single interaction involving a hard binary and a single may be sufficient to kick the binary away from the dense core on a higher orbit. In that case, the hardening of the binary would slow down.  Nevertheless, we have checked that the BBH sinking rate is higher than the binary--single interaction rate in the core by a factor of a few to several. Thus, we have neglected the effect of binary convection in the cluster.

The microphysics of the evaporation mechanism is as follows: during BBH--BH encounters, the BHs gain some kinetic energy and the hard BBHs harden (according to Heggie's law and as a consequence of the second law of thermodynamics). 
We have checked that ejections during a 3bb formation are unlikely, because typically the newly formed BBH is not hard enough to eject the third BH, and therefore we ignore them. 
In fact, Monte Carlo simulations of star clusters demonstrate that BH ejection during binary--single encounters is the primary way in which BHs are ejected from clusters~\cite{Weatherford:2022vbc}, together with the ejection of merger remnants that receive a relativistic kick. 

\subsection{Dynamical assembly channels of BBHs}
\label{sec:BBHformChannels}

Dense astrophysical systems are the sites of strong few-body interactions~\cite{1985IAUS..113..231H}.
We consider three channels of dynamical BBH assembly in clusters:
three-body interactions (Sec.~\ref{sec:ThreeBodyBinaryFormation}), two-body captures (Sec.~\ref{sec:Cap}), and binary--single exchanges (Sec.~\ref{sec:exchanges}).
In the following we describe all of these processes in detail, starting with a discussion of the original binary stars.

\subsubsection{Original binary stars}
\label{sec:OriginalBinaryStars}

Initially, we consider a fraction $f_{\rm b}$ of all stars to be in binary star systems.
The value of $f_{\rm b}$ has been observed to be as high as $30\%$~\cite{2012Sci...337..444S}, but it is an otherwise free parameter in our model.
Given the initial central luminous mass density $\rho_{\rm c,0}$, the number of binary stars per unit volume in the core of the cluster is $n_{\rm b}=f_{\rm b}\rho_{\rm c,0}/(2\overline{m})$.
For instance, fixing the central density of stars to be $10^5{M}_\odot{\rm pc}^{-3}$, we have $n_{\rm b}\simeq10^4~{\rm pc}^{-3}(f_{\rm b}/10\%)$ for $\overline{m}=0.6M_\odot$.
As for the binding energies of those binaries, we distribute them uniformly per unit logarithmic energy interval.
That is, the semimajor axes $a$ of binary stars are assumed to follow the log-flat distribution, $p(a)\propto1/a$, in the range from three solar radii (3R$_\odot$) up to a maximum value of $a_{\rm max}=0.1\times(4\pi\overline{m}\rho_{\rm c,0}/3)^{-1/3}$, which corresponds to the binary size being one tenth of the mean separation between stars in the core.
These choices are motivated by orbital distributions of binary stars in the solar neighborhood~\cite{1991A&A...248..485D}.
For simplicity, we neglect the contribution of original higher-multiplicity structures such as triples or quadruples (we do, however, take into account the dynamical assembly of BH triples formed via binary--binary interactions).
We also neglect the contribution of original BBH or BH--star pairs. Simulating their histories would require further modeling of astrophysical interactions in binary systems, such as mass trasfer and common envelope evolution.
We focus instead on dynamical formation, which should be the dominant BBH assembly channel in star clusters anyway.

Since we assume a log-uniform distribution for the semimajor axis of the original stellar binaries, a small minority of these original binaries will be soft and, hence, prone to disruption.
Thus we only keep track of the hard binaries.
If $a_{\rm h}=G\overline{m}/(4\sigma_\star^2)$ is the semimajor axis of a binary star at the hard-soft boundary, then the fraction of hard binaries in the cluster is given by $f_{\rm h}=\int_{3{\rm R}_\odot}^{a_{\rm h}}{\rm d}\ln a/\int_{3{\rm R}_\odot}^{a_{\rm max}}{\rm d}\ln a$.
Therefore, the number density of hard binary stars in the cluster is $f_{\rm h}n_{\rm b}$.

The effect of binary stars in the cluster is that they can exchange their stellar components in favor of heavy BHs, which can facilitate the formation of BBHs through a pair of successive exchanges: star--star$\to$BH--star$\to$BBH.
Since this chain of events occurs efficiently in the cluster core, the number of binary stars decreases with time.
To account for the dynamical formation of new hard binary stars, we also replenish their number through triple star interactions.
Ultimately the initial binary fraction $f_{\rm b}$ plays an important role in the dynamical assembly of BBHs from exchanges.
By default we take $f_{\rm b}=0.1$.

\subsubsection{Three-body binary formation}
\label{sec:ThreeBodyBinaryFormation}

The most common type of encounter in a dense cluster environment is the two-body interaction between two single objects.
However, due to energy conservation, two bodies with positive total energy approaching each other from infinity on hyperbolic orbits cannot bind, unless some amount of energy is dissipated away. We consider two mechanisms for energy dissipation: a third body carrying away the excess of energy (discussed in this section), and gravitational radiation during the approach (discussed in Sec.~\ref{sec:Cap} below).

Consider first the encounter of three single bodies that approach from infinity and interact in a small region. One of the bodies takes away enough energy to leave the other two bodies in a bound system, i.e., a newly formed BBH.
The probability that three bodies meet in a given region scales with the volume of that region, so this mechanism results into preferentially soft binaries.
Nevertheless, a fraction of these binaries will be hard enough to survive the long-term dynamics of the cluster, and will have a chance of merging in or out of the cluster.

The total three-body binary (3bb) formation rate in the dense core of a BH subsystem environment is derived in Appendix~\ref{app:ThreeBodyBinaryRate}. The expression we obtain there [Eq.~\eqref{Eq.Total_3bb_rate}] agrees with Eq.~(2) of Ref.~\cite{Morscher:2014doa} at the $\sim30\%$ level.
According to Ref.~\cite{1976A&A....53..259A} (see their Fig.~1), the probability of forming a binary during a strong three-body encounter is higher when the region of interaction is sufficiently small.
In fact, this probability asymptotically approaches $100\%$ when the size of the interaction volume is comparable to the semimajor axis of a binary that is marginally hard.
Therefore, in order to account only for those three-body interactions that induce hard binaries which survive, we assume $\eta\ge5$ by default, a choice consistent with Ref.~\cite{Morscher:2014doa}.
The mean timescale for the formation of a BBH somewhere in the core of the cluster due to three-body interactions among three single BHs is (assuming $\eta_{\rm min}=5$):
\begin{align}
\tau_{\rm 3bb}&\simeq 48.6\,{\rm Myr}\,\left({ \langle v_{\rm BH}^2 \rangle^{1/2} \over 10\,{\rm km\, s^{-1}} }\right)^9 \left({ \overline{m}_{\rm BH}\over20M_\odot }\right)^{-5}  \nonumber\\&\times  \left({ n_{\rm c,BH}\over 10^5\,{\rm pc}^{-3} }\right)^{-3} \left( {r_{\rm c,BH}\over0.1\,\rm pc} \right)^{-3}.
\label{Eq.3bbTime}
\end{align}%
Given that light objects cannot extract large amounts of energy when participating in a triple encounter, we neglect the formation of BBHs via BH--BH--star encounters (see Appendix~\ref{app:3bbWithStarAgent} for further details).

The strong dependence of this formula on $\langle v_{\rm BH}^2 \rangle^{1/2}$ and the mass is well known~\cite{1993ApJ...403..271G} and indicates that this channel is only important in environments with small velocity dispersion, i.e., in light clusters, which are ideal for the assembly of BBHs with heavy components from three-BH interactions~\cite{Franciolini:2022ewd}.
However, in the absence of hard binaries in massive clusters, there is no heat source in the center of the cluster causing its core to collapse. This leads to a dramatic increase in the central density of BHs, such that the velocity dependence can be compensated and 3bb formation becomes important. Since the relaxation timescale in the core is much smaller than $\tau_{\rm rh}$, this collapse occurs on a very small timescale. As such, the formation of hard binaries via the 3bb channel is inevitable even when the analytic formula in Eq.~\eqref{Eq.3bbTime} does not account for these fluctuations in the central BH density. We thus form a hard 3bb whenever the cluster is devoid of hard BBHs, and therefore see 3bbs forming even in massive clusters.

The 3bb rate formula $\Gamma_{\rm 3bb}(\eta'\ge\eta)$ (see Eq.~(2) from Ref.~\cite{Morscher:2014doa}) accounting for the formation of BBHs with hardness $\eta'$ larger than some threshold value $\eta$ is a complementary cumulative distribution function for $\eta$.
Thus, we draw the hardness ratio of a newly formed BBH from the derivative of this function, properly normalized. That is, the distribution for $\eta$ is derived from:
\begin{align}
	p(\eta){d}\eta=- {d\Gamma_{\rm 3bb}(\eta'\ge\eta)\over \Gamma_{\rm 3bb}(\eta'\ge5)  }.
\end{align}
Since $\eta\ge5$ for hard binaries, we approximate the $\eta$-dependence of the rate by writing $\Gamma_{\rm 3bb}\propto\eta^{-7/2}$, accounting only for gravitational focusing. Then we perform analytic inverse sampling to draw the hardness as $\eta=(1-u)^{-2/7}\eta_{\rm min}$, where $u$ is randomly drawn in the range $(0,1)$.
Once we get $\eta$, we use its definition, Eq.~\eqref{Eq:hardenessRatio}, to determine the semimajor axis.
The eccentricity of the new BBH is sampled from a thermal distribution, $p(e)de=de^2$.
Due to mass segregation, the most massive BHs sink to the central region and have a smaller velocity dispersion.
As a result, heavier BHs have a better chance of participating in few-body interactions in the core of the BH subsystem and of finding a binary companion.
When a new BBH forms, we account for these dynamical effects by sampling BH masses in a biased way.
We take the BH mass probability density function to be proportional to the rate.
For the 3bb channel, this gives $p_{\rm 3bb}(m_{\rm BH})\propto m_{\rm BH}^{5}$, where $m_{\rm BH}$ is the mass of each component of the newly formed BBH.

\subsubsection{Two-body captures}
\label{sec:Cap}

Here we consider the formation of BBHs through two-body interactions where the excess of energy is carried away by GWs. We refer to this process as two-body, or dynamical, captures. Being the most massive components in an evolved star cluster, BHs tend to sink to its core, where we can expect the capture process to play a role. Even there, BHs are rarely on hyperbolic orbits that are close enough to induce strong GW emission.
Turner~\cite{1977ApJ...216..610T} studied the emission of gravitational radiation from a system of two originally unbound point masses in the Newtonian limit (see also~\cite{1972PhRvD...5.1021H}).
If the amount of GW energy released during such a two-body encounter exceeds the total center-of-mass~energy of the system, a bound pair of BHs is formed.
The Keplerian orbital parameters of the newly captured BBH are given in Ref.~\cite{OLeary:2008myb} (see also Appendix~\ref{app:GravitationalCaptures}), with the eccentricity $e$ being extremely close to unity.
As such, the emission of gravitational radiation from such a binary is very efficient~\cite{Peters:1964zz}, and the pair merges after only a few orbits before it can be perturbed by a dynamical interaction with a third body, which could reduce the high eccentricity and increase the coalescence timescale~\cite{Jedamzik:2020ypm}.
In particular, the time required for a captured BBH to merge is estimated to be no larger than~\cite{OLeary:2008myb}
\begin{align}
\tau_{\rm me}^{\rm cap}\lesssim 1.6\,\mbox{kyr}\times{m_{\rm BBH}\over 30{M}_\odot}\left({\langle v_\infty^2\rangle^{1/2}\over10\,{\rm km\, s^{-1}}}\right)^{-3}.
\end{align}
The cross section for dynamical captures has been calculated in~\cite{1989ApJ...343..725Q,Mouri:2002mc} in the gravitational focusing regime.
From that we can find the timescale for a two-body capture between two equal-mass BHs to occur anywhere in the core of the cluster:
\begin{align}
	\tau_{\rm cap}&\simeq 154\,\mbox{Myr}\left({\overline{m}_{\rm BH}\over20{M}_\odot}\right)^{-2}\left({\langle v_\infty^2\rangle^{1/2}\over10\,{\rm km\, s^{-1}}}\right)^{{11\over7}} \nonumber\\ &\times  \left({ n_{\rm c,BH}\over 10^5\,\mbox{pc}^{-3} }\right)^{-2} \left({ r_{\rm c,BH}\over 0.1\,\mbox{pc}}\right)^{-3}.
\label{Eq.capTime}%
\end{align}%
This equation indicates that capture is more probable for two massive BHs, which experience a stronger gravitational focusing effect than lighter ones. 
In addition, the probability that a BH of mass $m_{\rm BH}$ will be captured by another BH of the same mass is $p_{\rm cap}(m_{\rm BH})\propto m_{\rm BH}^{2}$.
We use this probability density function to draw BH masses whenever a capture occurs in the simulation.
The dependence of $\tau_{\rm cap}$ on velocity is much weaker than for the 3bb channel [see~Eq.~\eqref{Eq.3bbTime}]. This implies that, as the cluster mass and hence velocity dispersion increases, capture dominates over BBH assembly via three-body interactions.

\subsubsection{Binary--single GW capture mergers}
\label{sec:BinarySingleGWcaptureMergers}

The outcome of an encounter between a hard binary and a single is again an outgoing binary and a single, unless the total energy is large enough to unbind the incoming binary.
During the resonant encounter the binary typically splits, and the binding energy oscillates among short-lived metastable binary states~\cite{Samsing:2013kua}.
So far we have discussed two outcomes at the end of these strong BBH-BH interactions: flybys and exchanges. 
Nevertheless, interactions involving BHs sometimes lead to the formation of highly eccentric intermediate states, resulting in the merger of their two components before the formation of the next state. 
These states are rare, however, due to ergodicity and to the frequency with which binary--single interactions occur.

A metastable BBH will undergo a GW capture merger during the resonant interaction if the energy emitted in the first pericenter passage exceeds the binding energy of the initial binary~\cite{Samsing:2017xmd}.
Equating the GW energy released in the parabolic limit (see e.g. Eq.~(22) from~\cite{Mouri:2002mc} with $e=1$) to the binding energy $Gm_1m_2/(2a)$, where $a$ is the semimajor axis of the initial binary, we obtain the critical maximum pericenter passage $\hat{r}_{p}$ required for a GW capture merger to occur between the $i$th and $j$th BH ($i,j=1,2,3$): 
\begin{align}
	\hat{r}_p={1\over2}\left({85\pi\over3\sqrt{2}}\right)^{2/7} \left[{\cal R}_s\left(  { (m_im_j)^{4/5}m_{ij}^{1/5} \over (m_1m_2)^{2/5} }  \right)\right]^{5\over 7}  a^{2/7},
\end{align}
where ${\cal R}_s(m)$ is the Schwarzschild radius of a BH with mass $m$ and $m_{ij}\equiv m_i+m_j$. Note that $m_3$ is the mass of the incoming single BH.

The pericenter $r_p$ and eccentricity $e$ are related by $r_p=(1-e)\tilde{a}$, where $\tilde{a}=m_im_ja/(m_1m_2)$ is the semimajor axis of the metastable state assuming binding energy conservation.
Thus, the necessary condition for a GW merger is $r_p<\hat{r}_{p}$, and it can be written as $e>\hat{e}$ in terms of the critical eccentricity that the metastable BBH should have to merge before a close interaction with the third body.
We further assume that all intermediate binaries have semimajor axis $\tilde{a}$, which is valid in order of magnitude~\cite{Samsing:2017xmd}.
If eccentricity is thermally distributed, we have $p(e>\hat{e})=1-\hat{e}^2\approx2(1-\hat{e})$ because the critical eccentricity must be close to unity, as discussed in Sec.~\ref{sec:Cap}.
Using the definition of pericenter distance, this can also be written as $p(e>\hat{e})=2\hat{r}_p/\tilde{a}$.
The probability for a GW capture merger to occur between two BHs during a single resonant BBH-BH encounter is given by $p_{\rm BBH-BH}^{\rm me}=1-[1-p(e>\hat{e})]^{N_{\rm IMS}}\approx p(e>\hat{e})N_{\rm IMS}$, where $N_{\rm IMS}$ is the number of intermediate BBH states and we assume that the probability $p(e>\hat{e})$ is small.

Numerical binary--single scattering experiments show that a typical resonant interaction goes through a few tens of intermediate BBH states.
Assuming an average of $N_{\rm IMS}=20$ (see Ref.~\cite{Samsing:2017xmd}, and the lower panel of Fig.~(6) in~\cite{Samsing:2013kua}) we have
\begin{subequations}
\begin{align}
	p_{\rm BBH-BH}^{\rm me}&=\left({85\pi\over3\sqrt{2}}\right)^{2/7}\left\{ {\cal R}_s\left( { (m_1m_2)^{3/5}m_{ij}^{1/5} \over (m_im_j)^{1/5} } \right)\over a \right\}^{5/7} N_{\rm IMS}\label{eq:BBHBHme}\\
	&\simeq2\cdot10^{-3}\times {N_{\rm IMS}\over20} \left({m\over10M_\odot}\right)^{5/7} \left({a\over1\,\rm AU}\right)^{-5/7},
\end{align}
\label{Eq.sb_merger_proba}%
\end{subequations}
where in the second line we have assumed that all BHs that participate in the interaction have the same mass $m=10M_\odot$.

Following a GW capture merger, since the binding energy has been lost in the form of GWs the merger product and third body typically do not form a bound system.
Moreover, given  the dependence of this probability on the masses, we sample the BHs that merge according to $p_{\rm BBH-BH}^{\rm me}(m_i, m_j)\propto[m_{ij}/(m_im_j)]^{1/5}$,
while the initial eccentricity of the merger is thermal in the range $[\hat{e},1)$.
In practice, we compute the three probabilities $p_{i+j}$ for a BBH--BH merger for each combination of BH pairs $i+j=1+2,2+3,3+1$ using Eq.~\eqref{eq:BBHBHme}. We then sample three pseudorandom numbers $w_i\in(0,1)$ and form the ratios $w_i/p_{i+j}$.
If the maximum ratio is greater than 1, then we perform the merger between the $i^{\rm th}$ and the $j^{\rm th}$ BH that gave the largest ratio.

\subsubsection{BBHs from exchanges in binary stars}
\label{sec:exchanges}

Original binary stars provide a reservoir of energy in the cluster that can be turned into binding energy of BBHs.
Binary--single exchanges provide a means for that energy transformation to happen and constitute yet another dynamical BBH formation channel.
It was previously shown that compact objects could form binaries with other stars via dynamical exchanges~\cite{1976MNRAS.175P...1H,Grindlay:2005ym}.
The substitution of a light member of a binary during its strong interaction with a third body becomes almost certain when the mass of the intruder significantly exceeds that of the companion~\cite{1980AJ.....85.1281H}.
The opposite trend is observed when the single is less massive than either of the binary companions. In that case the probability of exchange is very low, and the substitution rarely occurs.

Since BHs constitute the most massive objects in an evolved cluster, they tend to efficiently substitute stellar companions in stellar binaries, and very quickly become bound in binaries themselves.
These exchange processes become more efficient with increasing central density because of higher interaction rates~\cite{Miller:2008yw}.
In direct exchanges, in order to conserve the binding energy, the semimajor axis of the new binary must change by an amount $m_3/m_e$, where $m_3$ is the mass of the new member and $m_e$ the mass of the ejected companion~\cite{1996ApJ...467..359H,1993ApJ...415..631S}.
Since the mass of the intruder is usually larger, the size of the binary increases.
This in turn increases the cross section for subsequent encounters, and the stellar component of the newly formed BH--star binary, as the lighter companion, is likely to be substituted for another BH. 

The two successive exchanges $\mbox{star--star}\to\mbox{star--BH}\to\mbox{BBH}$, with the second process typically occurring on a shorter timescale, result in the formation of a wide BBH in the cluster.
Based on the fact that hardness is conserved during prompt exchanges, hard BBHs can form through this channel if the original binary star is hard to begin with.
To evaluate the exchange rates, we use a fitting formula for the cross section averaged over thermal eccentricities (Eq.~(19) from Ref.~\cite{1996ApJ...467..359H}), which agrees with numerical results to within 25\%:
\begin{widetext}
\begin{align}
\Sigma_{\rm ex}({1-2}\to{3-2})&\simeq14\,\mbox{AU}^2\times{a\over1\,\mbox{AU}} \left({ v_\infty\over10\,\mbox{km/s}}\right)^{-2}{m_{123}\over M_\odot}\left({m_{23}\over m_{123}}\right)^{1\over6}\left({m_3\over m_{13}}\right)^{7\over2}\left({m_{123}\over m_{12}}\right)^{1\over3}{m_{13}\over m_{123}}\nonumber\\ &\times \exp( 3.70 + 7.49x - 1.89y - 15.49x^2 - 2.93xy - 2.92y^2 + 3.07x^3 + 13.15x^2y - 5.23xy^2 + 3.12y^3).
\label{Eq:ExcCrossSection}
\end{align}
\end{widetext}
This equation provides the cross section for particle 1 of a hard binary $1$--$2$ with semimajor axis $a$ to be substituted for incoming single particle 3 with relative velocity at infinity $v_\infty$ to form a new pair $3$--$2$.
In addition, $m_{123}=m_1+m_2+m_3$ and $m_{ij}\equiv m_i+m_j$ with $i,j=1,2,3$ stand for the total mass of the system and pairwise sums of the masses, respectively, while $x=m_1/m_{12}$ and $y=m_3/m_{123}$.
Since the BBH formation channel in this case is a two-step process, we need to evaluate two timescales: $\tau_{\rm ex,1}$ for the first exchange $\mbox{star--star}\to\mbox{BH--star}$, and $\tau_{\rm ex,2}$ for the second exchange $\mbox{star--BH}\to\mbox{BBH}$.

Finally, we sample the masses of single BHs for the first exchange according to the probability density function $p(m_{\rm BH})\propto m_{\rm BH}$.
This stems from the approximate dependence of the product $\Sigma_{\rm ex}v_{\infty}$ on BH mass.
As an approximation we assume that an exchange takes place if the mass of the sampled BH is larger than the secondary component of the BBH, and if the sampled pericenter distance is such that $r_p<\max(m_1,m_2)a/(m_1+m_2)$.

\subsubsection{Binary-binary interactions}
\label{sec:BinaryBinary}

As discussed in the previous subsections, BBHs initially form with relatively wide orbits and readily interact with other objects. 
(Dynamical captures are an exception, as discussed in Sec.~\ref{sec:Cap}. The highly eccentric BBH that forms in that scenario, though possibly wide, merges promptly before any interaction can occur with a third object in the cluster.)
Since the binaries have larger cross sections, binary--binary interactions are also expected to play a significant role in the dynamics of the cluster. 
In particular, a BBH is likely to encounter another wide binary right after its formation when the semimajor axis is still relatively large ($\sim10^2-10^3$AU), so that it has not yet hardened significantly.
If the binary does not encounter another BBH soon enough, the probability for this to occur decreases with time, as the binary becomes tighter because of interactions with third singles and its cross section shrinks.

As mentioned in the introduction, we exploit the asymptotic theory of binary--binary interactions from Ref.~\cite{1980ApJ...241..618S} to approximate reactions rates and simplify the simulation.
Specifically, during the interaction of two binaries with a hierarchy in their binding energies, the harder of the two is treated as a single body which substitutes the lighter member of the wider pair.
To calculate the rate for this process, we use the binary--single exchange cross section, Eq.~\eqref{Eq:ExcCrossSection}, as well as other prescriptions outlined in Sec.~\ref{sec:exchanges}, with the tighter binary being treated as a single (a point mass).
Thus, the semimajor axis of the wider binary sets the pericenter scale of the interaction.

After the exchange occurs, a metastable triple system is formed.
Whether this three-body association persists as a hierarchical triple or not is determined by the stability condition explored in Ref.~\cite{2001MNRAS.321..398M}. The stability depends on the parameters of both the inner and outer~orbits, as well as on the angle between their angular momenta (the ``inclination'').
If the criterion for stability is not met, the metastable triple spontaneously breaks into its inner and outer components, which are no longer bound to each other.
Effectively, in that case we have a binary-binary interaction, in which the softer of the two BBHs dissociates and releases its components back into the cluster.
As a result of the breakup, the heaviest objects are most likely to find themselves in the surviving binary, because typically it is the lightest member that escapes to infinity~(see \cite{2006tbp..book.....V}, Chapter~7.4).
As the triple system dissociates, about $\varepsilon\approx38\%$ of the binding energy of the third distant body is transferred to the inner binary~\cite{Zevin:2018kzq}. 
This leads to tightening of the harder binary, whose new semimajor axis (by energy conservation) is
\begin{align}
{a_1'\over a_1}={1\over1+\varepsilon{m_3m_4\over m_1m_2}{a_2\over a_1}}\,,
\end{align}
where $a_2>a_1$ is the original semimajor axis of the binary which breaks.
We also resample the thermal eccentricity of the surviving binary to model random perturbations in the orbital angular momentum~\cite{1937AZh....14..207A}.

\renewcommand{\arraystretch}{1.4}
\setlength{\tabcolsep}{5pt}
\begin{table*}
\centering
\caption{List of all input parameters of our code (first column) along with a short description (third column). The second column shows the option to set while executing the program in a command line interface. If a variable is not passed upon program call, then the default value listed in the last column is assumed.}
\vspace{.1cm}
\begin{tabular}{c c c c}
\toprule
Input &  
Flag & 
Description &
Default value
\\
\midrule

$N_{0}$ & -N & Initial number of stars &  $10^6$ \\

$r_{\rm h,0}$ & -r & Initial half-mass radius [pc] & 1\,pc \\

$m_{\rm ZAMS}^{\rm min}$ & -mm & Minimum ZAMS star mass [$M_\odot$] & 0.08$M_\odot$ \\

$m_{\rm ZAMS}^{\rm max}\in(20,340]$ & -mM & Maximum ZAMS star mass [$M_\odot$] & 150$M_\odot$ \\

$Z\in[10^{-4},0.02]$ & -Z & Absolute metallicity & 0.001 \\

$z_0$ & -z & Redshift of cluster formation & 3\\

$n_{\rm star,0}$ & -n & Initial central stellar number density & $10^6$  \\

$f_{\rm b}\in[0,1]$ & -fb & Initial binary star fraction & $10\%$ \\

$-$ & -S & Random number generator seed & 1234567890 \\

$dt_{\rm min}$ & -dtm & Minimum simulation timestep [Myr] & $0.1$\,Myr \\

$dt_{\rm max}>dt_{\rm min}$ & -dtM & Maximum simulation timestep [Myr] & 50\,Myr \\

$t_{\rm max}$ & -tM & Maximum simulation time [Myr] & $14,000$\,Myr \\

$\sigma_{\rm kick}$ & -wK & One-dimensional natal velocity kick dispersion of BHs [$\rm km\, s^{-1}$] & $265\,\rm km\, s^{-1}$ \\

$-$ & -K & Natal kick prescription (0 for fallback, 1 for momentum conservation) & 1 \\

$R_{\rm gal,0}$ & -R & Initial galactocentric radius [kpc] & 8\,kpc \\

$v_{\rm gal}$ & -vg & Galactocentric circular velocity [$\rm km\, s^{-1}$] & $220\,\rm km\, s^{-1}$ \\

$0\le\chi_{\rm 1g}<1$ & -s & Natal spin parameter of first generation (1g) BHs & 0\\

$-$ & -SD & Natal spin distribution ($0$ for uniform, $1$ for monochromatic) & 0\\

$-$ & -P & Print runtime information (0 for no, 1 for yes) & 1 \\

$-$ & -Mi & Export mergers file (0 for no, 1 for yes) & 1 \\

$-$ & -MF & Name of .txt output file with BBH merger source parameters & ``mergers'' \\

$-$ & -Ei & Export evolution file (0 for no, 1 for yes) & 1 \\

$-$ & -EF & Name of .txt output file with time-dependent quantities & ``evolution'' \\

$-$ & -Hi & Export hardening file (0 for no, 1 for yes) & 1 \\

$-$ & -HF & Name of .txt output file with BBH time evolution information & ``hardening'' \\

$-$ & -BIi & Use external BH file (0 for no, 1 for yes) & 0 \\

$-$ & -BIF & Name of .npz input file with initial BH masses & ``input$\_$BHs.npz'' \\

$-$ & -BOi & Export BH masses file (0 for no, 1 for yes) & 1 \\

$-$ & -BOF & Name of .npz file with the masses of all BHs in $M_\odot$ & ``output$\_$BHs.npz'' \\

\bottomrule
\end{tabular}
\label{Tb:parameters}
\end{table*}

If a hierarchical triple forms through a binary--binary interaction, the inner binary may experience a von Zeipel-Lidov-Kozai (ZLK) excitation~\cite{1910AN....183..345V,1962P&SS....9..719L,1962AJ.....67..591K,2019MEEP....7....1I} contingent upon the orientation and orbital elements of the inner and outer orbits.
As a consequence of angular momentum exchange between the inner and outer binaries, the eccentricity $e_1$ of the inner pair may be driven to values close to unity in an oscillatory fashion. During the oscillations, eccentricity is traded off for relative inclination $\iota$, so that the quantity $\sqrt{1-e_1^2}\cos\iota$ is constant at first order in the perturbation.
If the inner pair is a BBH (as assumed here), GW emission becomes efficient only near the maximum of the ZLK cycle, where the eccentricity is the highest.
If this maximum eccentricity is close to unity, the inner pair hardens quickly and can merge because of GW emission over several ZLK cycles.
As discussed in Ref.~\cite{Miller:2002pg}, interaction of the triple with a fourth single object tends to alter the parameters of the triple orbits, most notably the orientations, and thus interfere with the secular evolution of the triple.
However, if the ZLK timescale is short enough, the inner BBH may merge before the next interaction of the triple with the single perturber.
It is also possible that the orientations of the triple's orbits are initially not optimal for an efficient ZLK excitation, but an encounter of the triple with a single star perturbs the triple's parameters in such a way that the inner pair merges via the ZLK channel.

To account for secular evolution of the triple over many ZLK oscillation cycles, we assume that the orbit of the inner pair with components $m_0$ and $m_1$ and semimajor axis $a_1$ decays due to GW emission every time it passes through its maximum eccentricity phase $e_{\rm max}$. Then the time it takes for it to merge via the ZLK excitation channel in the high-eccentricity regime reads~\cite{Miller:2002pg}
\begin{align}
\tau_{\rm me}^{\rm ZLK}(e_{\rm max}\to1^{-})&\simeq2.5\times10^5\,\mbox{Gyr}\,{2000M_\odot^3\over (m_0+m_1)m_0m_1}\nonumber\\&\times\left({a_1\over1\,\mbox{AU}}\right)^4(1-e_{\rm max}^2)^3,
\end{align}
which contains an extra factor of $\sqrt{1-e_{\rm max}^2}$ compared to Peters' coalescence timescale~\cite{Peters:1963ux,Peters:1964zz} in the high eccentricity regime~\cite{1997AJ....113.1915I}.
This extra factor accounts for the acceleration of the inspiral and the emission of GWs, which is dominant during the time spent near the maximum eccentricity phase of the ZLK cycle.
The condition that the inner pair merges before the next triple-single encounter is $\tau_{\rm me}^{\rm ZLK}<\tau_{\rm enc}$, where $\tau_{\rm enc}$ is given by Eq.~\eqref{Eq:CollTime}.
Most of the triples that do not satisfy this condition do not merge, and rather get disrupted after interacting with a star.

We approximate a triple-single encounter as a binary--single interaction with the outer pair playing the role of the binary (as it has a larger cross section), and calculate binary-binary interaction timescales $\tau_{bb}$ using Eq.~\eqref{Eq:CollTime}.
Since it is more probable that the single is a typical star, the outer pair of the triple hardens, reducing its semimajor axis while slightly increasing its eccentricity.
Here it suffices to consider only triples where the tertiary is strongly bound to the inner pair, such that the outer orbit is a hard binary. This is justified by the fact that the progenitor binaries that gave rise to the triple were hard, and by conservation of energy during the exchange.

The triple persists in the cluster as long as the stability criterion is met. This criterion breaks down if the tertiary comes dangerously close to the inner orbit, which can render the triple unstable so that it splits into a binary and a single unbound BH, with the least massive component having the highest probability of escaping after the resonant breakup.

Finally, we also calculate the BH-star--BH-star collision timescale $\tau_{pp}$ at every step of our simulation.
The outcome of such an interaction is taken to be the formation of a BBH, with its eccentricity being thermal and its semimajor axis determined by energy conservation (and assuming that the two stars do not form a bound system after the interaction).
Note that we do not evolve the orbital parameters of BH--star binaries. Rather, we just keep track of their current number in the cluster and of their properties, and use this as an intermediate population which can produce BBHs through either BH-star--BH or BH-star--BH-star interactions.
We neglect the effect of triple--binary interactions due to the rarity of these events.

\subsection{Our simulation algorithm}
\label{sec:SimulationDetails}

\subsubsection{Code input parameters}
\label{sec:CodeParameters}

Table~\ref{Tb:parameters} lists the input parameters of our code. The first column summarizes the notation for the parameters adopted in this paper, while the second column gives the respective flags to be passed to the code in a command line interface. A brief description of each parameter is provided in the third column.
If the value of a parameter is not specified, the code will run with the default value listed in the last column of the table. More information on how to run the code, along with requirements and output files, can be found at the URL \url{https://github.com/Kkritos/Rapster/blob/main/README.md}.

\subsubsection{Global simulation timestep}
\label{sec:GlobalSimulationDetails}

Given the many processes that can occur at any time in our simulations, we pick the timestep $dt_1$ of the global iteration to be controlled by the minimum timescale for BBH formation.
For computational efficiency and sufficient time resolution, we do not allow the timestep to be smaller or larger than some user-defined threshold values. If not defined by the user, these are set to the default values $dt_{\rm min}=0.1\,\rm Myr$ and $dt_{\rm max}=50\,\rm Myr$, respectively (see~Table~\ref{Tb:parameters}).
Moreover, to avoid numerical artifacts, we do not allow $dt_1>t_1$ whenever $t_1>0$.
Thus, the timestep of our global simulation is adaptive, and given by
\begin{align}
dt_1 =\min\Big\{ &t, dt_{\rm max}, \nonumber \\ & \max\big[ dt_{\rm min},  \min(\tau_{\rm 3bb},\tau_{\rm cap},\tau_{\rm ex,1},\tau_{\rm ex,2})\big]\Big\}.
\label{Eq.global_timestep}
\end{align}
We remind the reader that $\tau_{\rm ex,1}$ and $\tau_{\rm ex,2}$ correspond to the timescales for star-star$\to$BH-star and BH-star$\to$BBH exchanges, respectively.
In addition, if $t<\tau_{\rm cc}$ we set $dt_1=dt_{\rm min}$.

After setting the global timestep, we sample the number of occurences $k_i$ for each process $i$ from the Poisson distribution with parameter $dt_1/\tau_i$.
Here $i$ corresponds to one of the channels, (i.e., $i=\mbox{3bb}$, cap, ex,1, ex,2, bb, pp).
For this step, we create $k_i$ binaries following the prescriptions for each channel described above. 
If $k_i=0$ for every channel and $t_1>\tau_{\rm cc}$, no process occurs and the simulation proceeds to the next stage, which is to evolve the cluster's properties (mass, radius, and galactocentric radius) as well as BBHs and BH triples (if present in the cluster).
By virtue of Eq.~\eqref{Eq.global_timestep}, in this case we have $dt_1=\min(t_1,dt_{\rm max})$.
We evolve every single BBH in a separate routine.
We describe the details of this local routine in Sec.~\ref{sec:BBHevolAlgorithm}.

\begin{figure}
\centering
\includegraphics[width=0.3\textwidth]{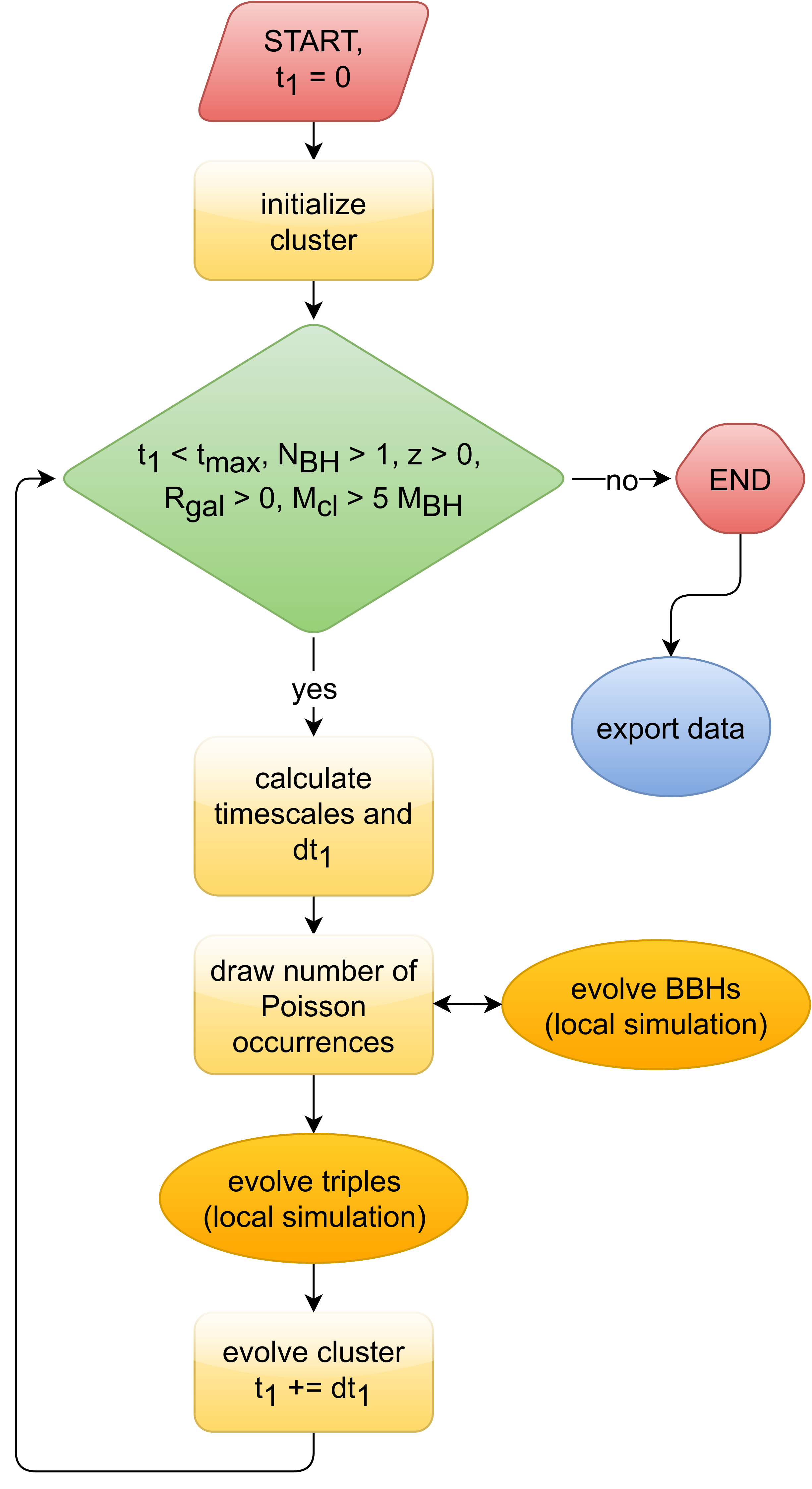}
\caption{Flowchart summarizing our global simulation algorithm. In this figure, yellow rectangles represent dynamical processes, and the green rhombuses are if-statement conditions. The blue ovals correspond to either termination or continuation of the loop, while the red shapes define the starting and ending points. 
The orange ovals have a similar meaning to the yellow rectangles, but encompass a more sophisticated algorithm (see Fig.~\ref{Fig.BBHevol_flowchart} for the ``evolve BBHs'' block).
Arrows show the flow of the algorithm. 
This flowchart was made with~\href{https://www.diagrams.net}{\tt draw.io}.}
\label{Fig:globalSimulation}
\end{figure}

Figure~\ref{Fig:globalSimulation} represents a flowchart that summarizes the global algorithm.
The simulation continues to loop as long as all the following termination conditions are satisfied:
\begin{itemize}
\item the current global simulation time $t_1$ does not exceed a predetermined maximum value $t_{\rm max}$ (set to $14,000\,\rm Myr$ by default);
\item there is more than one single BH retained in the cluster at the current step (enumerating members of binaries in the core);
\item redshift $z=0$ is not reached;
\item the cluster's galactocentric radius is positive;
\item the cluster mass is $5$~times larger than the total mass in BHs retained in the cluster.
\end{itemize}
The last condition is there to ensure that the total mass in BHs is relatively small, so that the approximations we use are valid.
The factor of 5 was chosen arbitrarily and is safe, considering that under the standard IMF the total BH mass fraction does not exceed 10\% (initially).
By default, we evolve the system for at most a Hubble time $\simeq13.8\,\rm Gyr$. Most simulations terminate sooner than that, either because the cluster has evaporated, because the number of BHs has been depleted, or because redshift zero has been reached.

\begin{figure*}
\centering
\includegraphics[width=0.8\textwidth]{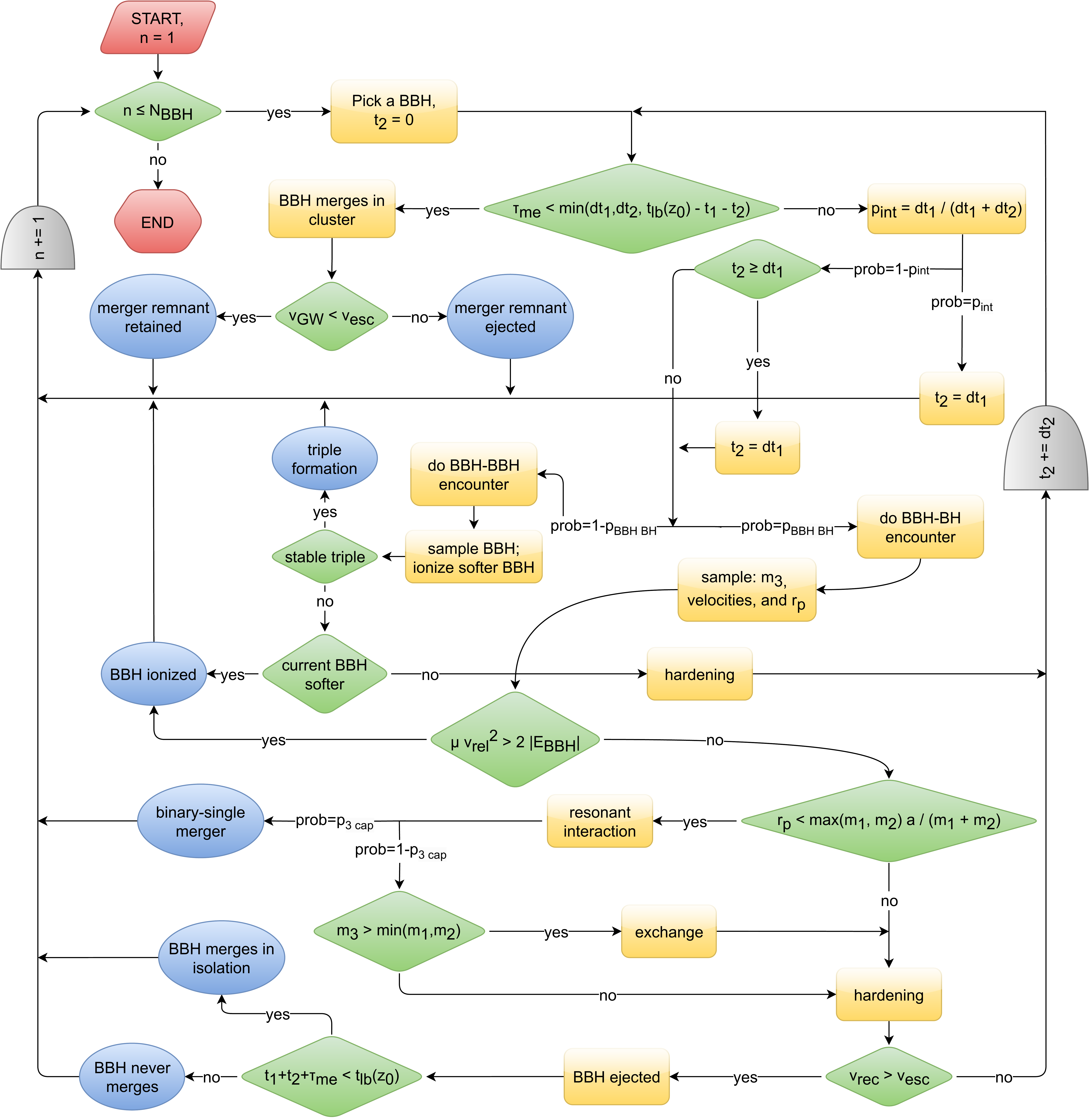}
\caption{Sketch of the flowchart of our local algorithm to evolve single BBHs in the cluster (see Sec.~\ref{sec:BBHevolAlgorithm}). A description of the different shapes in this figure can be found in the caption of Fig.~\ref{Fig:globalSimulation}. Gray semidisks denote logical conjunctions with update commands. A description of all variables that appear in this figure follows: ``N$_{\rm BBH}$'' is the current number of BBHs; ``n,'' the local simulation iteration index; ``m$_1$'' and ``m$_2$,'' the component masses of the current BBH, ``a'' its semimajor axis, and ``|E$_{\rm BBH}$|'' its binding energy; ``t$_1$,'' ``t$_2$'' and ``dt$_1$,'' ``dt$_2$'' the global and local times and timesteps, respectively; ``t$_{\rm lb}$(z$_0$)'' the lookback time at the redshift of cluster formation ``z$_0$''; ``$\tau_{\rm me}$,'' the BBH GW merger time; ``v$_{\rm GW}$,'' ``v$_{\rm rel}$,'' ``v$_{\rm rec}$,'' ``v$_{\rm esc}$'' are, in order, the GW merger kick, preencounter relative velocity, post-encounter binary recoil, and escape speed; ``p$_{\rm int}$,'' the probability for binary--single interaction happening within the global timestep; ``p$_{\rm 3\ cap}$,'' the probability for binary--single GW capture merger; ``m$_3$'' the third mass and ``$\mu$'' the reduced mass of the binary--single system; finally, ``r$_{\rm p}$'' is the pericenter distance of the interaction. This flowchart was made with~\href{https://www.diagrams.net}{\tt draw.io}.}
\label{Fig.BBHevol_flowchart}
\end{figure*}

\subsubsection{Local BBH evolution algorithm}
\label{sec:BBHevolAlgorithm}

Figure~\ref{Fig.BBHevol_flowchart} shows a flowchart of the local routine used to evolve single BBHs while they are inside the cluster.
At every step of the global simulation (at the moment of global simulation time $t_1$, with the current timestep being set to $dt_1$), we run the routine for each single BBH in the cluster.
This routine is characterized by local time $t_2$, which is set to zero whenever the local simulation starts for a BBH.
To avoid any biases, we first randomize the order of the BBHs in the array of binaries.
We evolve the BBH by determining whether it interacts with a single BH or not based on the probability for interaction within a global timestep, while the corresponding BBH--BH interaction time sets the local timestep $dt_2$.
That probability is computed as follows: $p_{\rm int}=dt_1/(dt_1+dt_2)$.

Whether a BBH interacts with a single BH or another BBH is a random process depending on the number densities of these two populations.
These interactions occur with the corresponding probabilities, which we calculate as
\begin{subequations}
	\begin{align}
		p_{\rm BBH-BH} &= {\tau_{\rm BBH-BBH} \over \tau_{\rm BBH-BH} + \tau_{\rm BBH-BBH} },\\
		p_{\rm BBH-BBH}  &= 1 - p_{\rm BBH-BH},
	\end{align}
	\label{Eq.ProbaInter}%
\end{subequations}
where $\tau_{\rm BBH-BH}$ and $\tau_{\rm BBH-BH}$  are the timescales over which the BBH encounters a single BH or a BBH, respectively, and are calculated using Eq.~\eqref{Eq:CollTime}.
If the number density of BHs in the core is higher than that of BBHs, then $\tau_{\rm BBH-BH}<\tau_{\rm BBH-BBH}$, and the probability that the BBH interacts with a BH is higher.
Even though there is only a handful of binaries at any time during the simulation, their cross section tends to be higher, because their BHs are well separated and they tend to be heavier. As such, binary--binary interactions are frequent in the core, and a fraction of them leads to the formation of hierarchical triples.

Depending on whether the binary is tight enough to merge in the cluster (condition ``$\tau_{\rm me}<\min(dt_1,dt_2,t_{\rm lb}(z_0)-t_1-t_2)$'' in the chart) or the encounter energetic enough to ionize it (condition ``$\mu v_{\rm rel}^2>2|E_{\rm BBH}|$''), we respectively let the binary coalesce or dissociate, releasing its components as single BHs back into the cluster.

If the BBH merges within the cluster, we check how the GW recoil compares with the escape velocity, and determine whether the remnant will be ejected.
When that happens, the single BH merger remnant lives in the low-density field, and cannot participate in dynamical processes leading to the formation of higher-generation BHs.
To be consistent, we also account for the self-gravity of the BHs when estimating the escape velocity as $v_{\rm esc}=2\sqrt{\langle v_\star^2 \rangle + \langle v_{\rm BH}^2 \rangle}$, although this contributes only a small correction.
If the recoil velocity lies within the range from $2\langle v_{\rm BH}^2\rangle^{1/2}$ to $v_{\rm esc}$ then the binary is not ejected from the cluster, but rather it escapes the BH subsystem and moves to a higher orbit. We then add a delay to the interaction rate that corresponds to the dynamical friction timescale required for the binary to return to the BH core. This convection timescale is given by $(\overline{m}/m_{12})\tau_{\rm rh}$, where $\tau_{\rm rh}$ is the current half-mass relaxation time and $m_{12}$ the mass of the convecting BBH.
Moreover, if the binary does not merge or is not disrupted, we evolve its orbital parameters according to Eq.~\eqref{Eq.HardeningEq}.
If the local simulation time does not exceed the global timestep (condition ``$t_2\ge dt_1$'' in the chart) and the binary is not ejected from the cluster or does not happen to interact with another BBH (condition ``$u<p_{\rm int}$''), we update the local iteration timestep (``$t_2+=dt_2$'') and proceed to consider the next collision for the same binary.
After we have an outcome for the BBH (or the local timestep is exceeded) we move on to run the local routine for a different BBH, and repeat the scheme until all BBHs have been evolved.

\section{Comparison with the Cluster Monte Carlo}
\label{sec:Comparison_with_CMC}

In this section we test the physical predictions of {\tt Rapster} by carrying out direct comparisons with the Cluster Monte Carlo ({\tt CMC}) code~\cite{Rodriguez:2021qhl}.
The {\tt CMC} is a publicly released indirect $N$-body code that simulates dense stellar systems implementing a H\'enon-type orbit-averaging technique for collisional dynamics.

The physical quantities we compare include the time evolution of cluster mass, the half-mass radius, the number of BHs, and BBH merger properties such as their mass distribution, merger times, the branching ratio of different merger channels, and the orbital parameters of ejected pairs. 
All of these comparisons will be carried out on a $4\times4$ grid of {\tt CMC} models from the public {\tt CMC} cluster catalog presented by Kremer {\it et al.}~\cite{Kremer:2019iul}. 

The {\tt CMC} models we compare against are characterized by their number of particles $N$, virial radius $r_{\rm v}$, galactocentric radius $R_{\rm gal}$, and metallicity $Z$.
We fix $R_{\rm gal}=20\, \rm kpc$, a galactocentric velocity of $220\, \rm km\, s^{-1}$, and $Z=0.002$, while we vary $N\in\{2,4,8,16\}\times10^5$ and $r_{\rm v}\in\{0.5,1,2,4\}\, \rm pc$.
We choose a large value for the galactocentric radius because we would like to ignore the external tidal effects of the galaxy as much as possible, and concentrate on the internal dynamical evolution.
Using the definition of virial radius and the virial theorem, we set the initial half-mass radius to be $r_{\rm h}=0.8r_{\rm v}$.
The initial binary fraction is set to $f_b=5\%$, as in Ref.~\cite{Kremer:2019iul}.
In our model we do not incorporate isolated binary evolution. Therefore, a nonzero initial binary fraction will only affect the efficiency of the exchange formation channel: star--star$\to$BH--star$\to$BH--BH.

Kremer {\it et al.}~\cite{Kremer:2019iul} have an initial distribution of stars that follows the King profile with parameter $W_0=5$.
According to Table~II of Ref.~\cite{1966AJ.....71...64K}, this corresponds to a concentration parameter of $c=10.70$ and to a parameter $\mu=11.81$.
From Eq.~(40) of~\cite{1966AJ.....71...64K} we can relate the stellar number $N$ to the core radius $r_{\rm c}$ and central number density $n_0$ as $N=n_0r_{\rm c}^3\mu$, assuming there is no primordial mass segregation.  Then, the integral of the King expression for the stellar number density $n(r)4\pi r^2dr$ (where $n(r)$ follows Eq.~(14) of~\cite{1962AJ.....67..471K}) from 0 to the half-mass radius $r_{\rm h}$ yields $N/2$, and this returns an algebraic equation that relates $r_{\rm c}$ to the other variables. 
Defining $x\equiv r_{\rm h}/r_{\rm c}$ and combining with $N=n_0r_{\rm c}^3\mu$, the numerical solution to the equation above is found to be $x\simeq1.539$. This allows us to write the initial stellar central density, which is one of the input parameters of {\tt Rapster}, as $n_0=Nx^3/(r_{\rm h}^3\mu)$.

\begin{figure}
	\centering
	\includegraphics[width=0.5\textwidth]{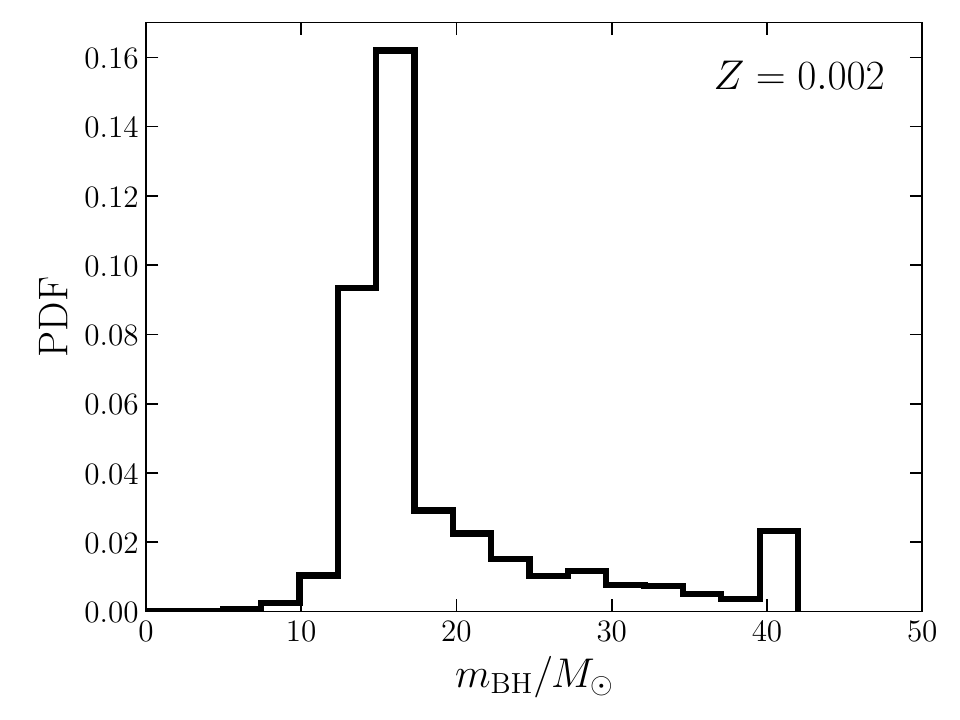}
	\caption{Initial BH mass spectrum used in our simulations. This mass spectrum is estimated from the $100\,\rm Myr$ snapshot of the $r_{\rm v}=4\,\rm pc$ cluster models, as described in the main text (Sec.~\ref{sec:Comparison_with_CMC}). We set the absolute metallicity to be $Z=0.002$.}
	\label{Fig:BH_spectra}
\end{figure}

To make sure that the initial conditions are as similar as possible, we input the BH masses and number of BHs from the first {\tt CMC} snapshot at about $100\,\rm Myr$.
However the BH mass spectrum varies with time. Systems with smaller $r_{\rm v}$ have a small relaxation time, and thus evolve faster.
This is evident especially in the $r_{\rm v}=0.5\,\rm pc$ models, for which most heavy BHs in the pair-instability pileup are ejected from the cluster within $100\,\rm Myr$.
The shape of the initial BH mass spectrum primarily depends on the assumed stellar physics.
The BHs in {\tt CMC} are generated assuming a Kroupa IMF in the range $[0.08,150]M_\odot$ and using the rapid Fryer et al. (2012) remnant-mass prescription with mass fallback~\cite{Fryer:2011cx}.
Moreover, all first-generation BHs are initially given a zero natal spin.
To ensure that the BH mass spectrum at $100\,\rm Myr$ is as close to the initial {\tt CMC} mass spectrum as possible (but not precisely the same), we take the spectrum of the first {\tt CMC} snapshot models with $r_{\rm v}=4\,\rm pc$, because the relaxation timescale is large enough for very little BH ejection to occur (excluding natal kicks).
Then we set up our initial conditions by stacking all BH masses of the first snapshot from all four $r_{\rm v}=4\,\rm pc$ models together to improve the statistics, and by sampling BHs with repetition.
Moreover, we discard any BH mass above the BH high-mass pileup, because those would correspond to second-generation products retained in the cluster.
Indeed, at least a few mergers do occur within $100\,\rm Myr$, while by construction the BH mass does not exceed $\sim40M_\odot$ (see Ref.~\cite{Kremer:2019iul}, Fig.~1).
In Fig.~\ref{Fig:BH_spectra} we show the resulting BH mass distribution used as input for our simulations.
Most BHs below $10M_\odot$ receive a large enough natal kick to be ejected from the cluster at formation.

In the following we discuss in detail our results and compare them to single {\tt CMC} realizations, focusing first on the global evolution of the star cluster (Sec.~\ref{Sec:Global_evolution}) and then on the properties of dynamically formed BBH mergers (Sec.~\ref{Sec:Binary_black_hole_merger_properties}).

\subsection{Global evolution of the star cluster}
\label{Sec:Global_evolution}

The global evolution of the star cluster (in particular, the time evolution of its total mass and half-mass radius) strongly affects the evolution of its BH subsystem. 
Many quantities characterizing the subcluster, such as the half-mass radius and core radius of the BH subsystem [cf.~Eq.~\eqref{Eq.segRadius} and~\eqref{Eq.BHcoreRadius}], depend directly or indirectly on the properties of the whole system. In turn, these quantities control the central BH density (and thus the merger rates) as well as the escape velocity, which is linked to the BH ejection rate.
As a reminder for the reader, the method we use to evolve the star cluster is described in Sec.~\ref{sec:Star_cluster_evolution} and Sec.~\ref{sec:GlobalSimulationDetails}.

\begin{figure*}
	\centering
	\includegraphics[width=1\textwidth]{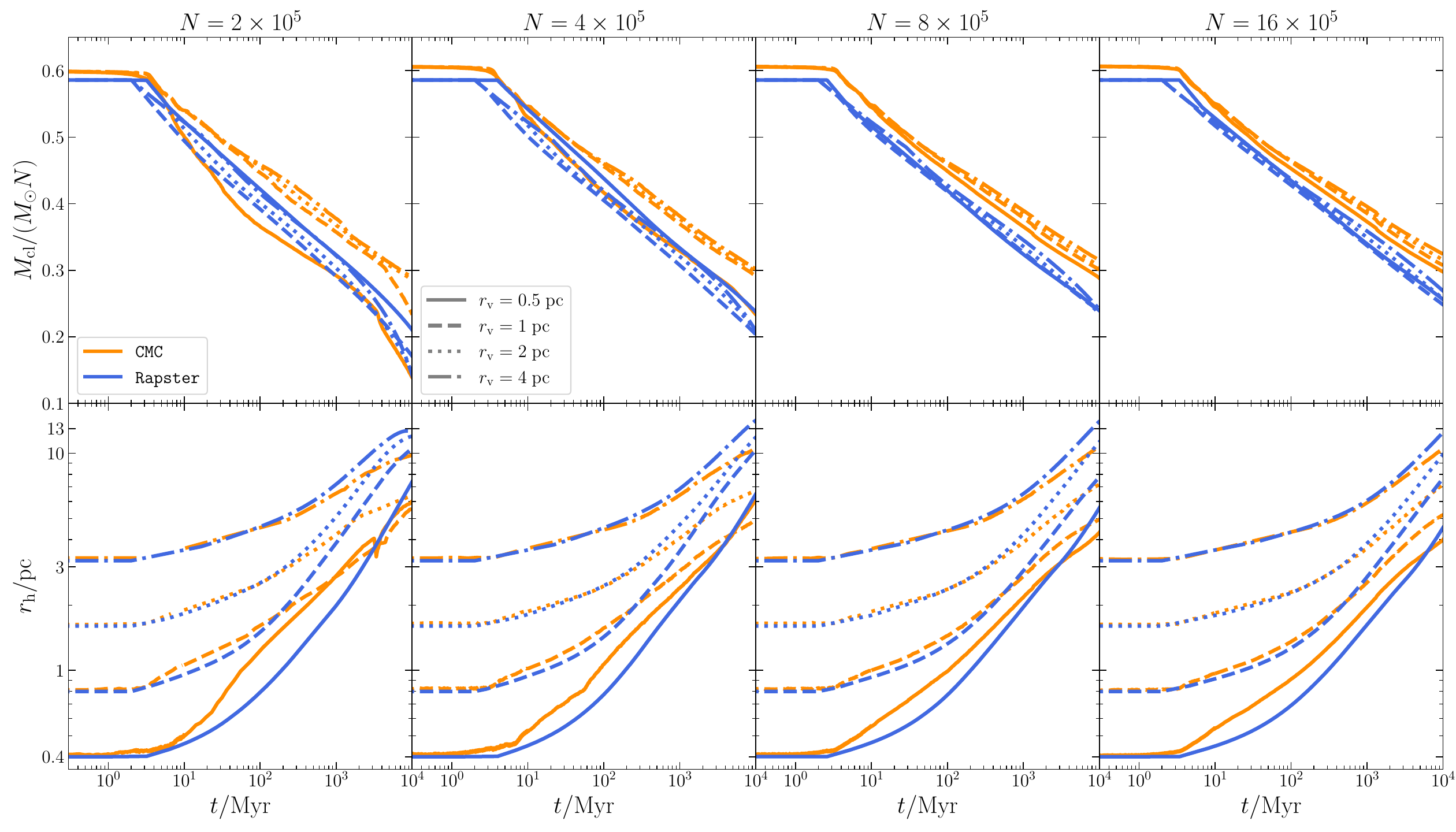}
	\caption{Single realizations of the time evolution of the total normalized star cluster mass (top panels) and of the half-mass radius (bottom panels) for a grid of cluster models with initial stellar number $N\in\{2,4,8,16\}\times10^5$ (left to right) and initial virial radii $r_{\rm v}\in\{0.5, 1, 2, 4\}\, \rm pc$, as indicated in the legend. The galactocentric radius is $R_{\rm gal}=20\, \rm kpc$ and the absolute metallicity is $Z=0.002$ for all simulations shown in this plot. The {\tt CMC} data are taken from Ref.~\cite{Kremer:2019iul}.}
	\label{Fig:global_evolution}
\end{figure*}

In Fig.~\ref{Fig:global_evolution} we show the evolution of the total cluster mass $M_{\rm cl}$ (top panels) and half-mass radius $r_{\rm h}$ (bottom panels) for the grid of simulations discussed at the beginning of this section.
Different columns refer to a different initial number of stars $N$.
The plot focuses on specific random realizations of the {\tt CMC} models, and our results are directly compared against the {\tt CMC} data. 

\subsubsection{Cluster mass}

As can be seen in the top panels of Fig.~\ref{Fig:global_evolution}, the cluster mass evolution is weakly dependent on the initial virial radius of the cluster. 
Clusters start losing mass after about $3\,\rm Myr$, at which point the evolution of massive stars dominates the evolution of the cluster.
At later times the clusters lose stars with velocity above the escape velocity as a consequence of relaxation processes.
Since BHs represent a small fraction of the total cluster mass, BH ejections do not directly affect $M_{\rm cl}$.
The very small differences at early times are due to the slightly different average cluster mass, but they are not significant.
In {\tt Rapster} the average mass is estimated to be $\simeq0.58M_\odot$ using the Kroupa IMF~\cite{Kroupa:2000iv} in the range $[0.08M_\odot,150M_\odot]$ with spectral values $a_0=-0.3$, $a_1=-1.3$, $a_2=-2.3$, and $a_3=-2.3$ and mass boundaries $m_1=0.08M_\odot$, $m_2=0.50M_\odot$, and $m_3=1M_\odot$.

In reality, depending on the star formation efficiency, the cluster contains some residual gas whose expulsion would affect the evolution of the cluster in the first million years, as shown in Ref.~\cite{Baumgardt:2007av}. However, the simulations discussed in this paper refer to dry systems (i.e., we ignore the effect of gas).

\subsubsection{Half-mass radius}
\label{Sec:Half_mass_radius}

The cluster expands as a consequence of energy production in its core, primarily from the formation of hard BBHs and their subsequent hardening, which heats up the system following the initial core collapse.
The results in the lower panels of Fig.~\ref{Fig:global_evolution} imply that the expansion of the cluster is not very strongly correlated with the particle number $N$. In general, initially more compact clusters end up being slightly more compact than their initially sparser counterparts.
Nevertheless, the smaller $r_{\rm v}$ is, the larger the factor by which the cluster expands. This is because more compact systems have higher binary formation rates, which leads to a larger energy output from the core.

When compared to their {\tt CMC} counterparts, {\tt Rapster} models consistently expand to relatively larger values of $r_{\rm h}$ at late times (after a few billion years of evolution).
This is related to the BH ejection rate. The relaxation term in the differential equation for $r_{\rm h}$ in Eq.~\eqref{eq:half_mass_radius_evolution_equation} is proportional to $\tau_{\rm rh}^{-1}$, which is directly proportional to the multimass factor $\psi$. We recall that in the case of two-mass models (assumed in {\tt Rapster} as an approximation) we have $\psi=1+S$, where $S$ is the Spitzer factor.
If the ejection rate of BHs is reduced, then $S$ decreases at a slower rate and thus the expansion rate of the cluster is larger, since $dr_{\rm h}^{\rm (rlx)}/dt\propto(1+S)$.
In the {\tt Rapster} simulations we observe a slightly slower BH ejection rate, particularly for lower-mass systems.

\subsubsection{BH subsystem evaporation}
\label{Sec:BH_subsystem_evaporation}

\begin{figure*}
	\centering
	\includegraphics[width=1\textwidth]{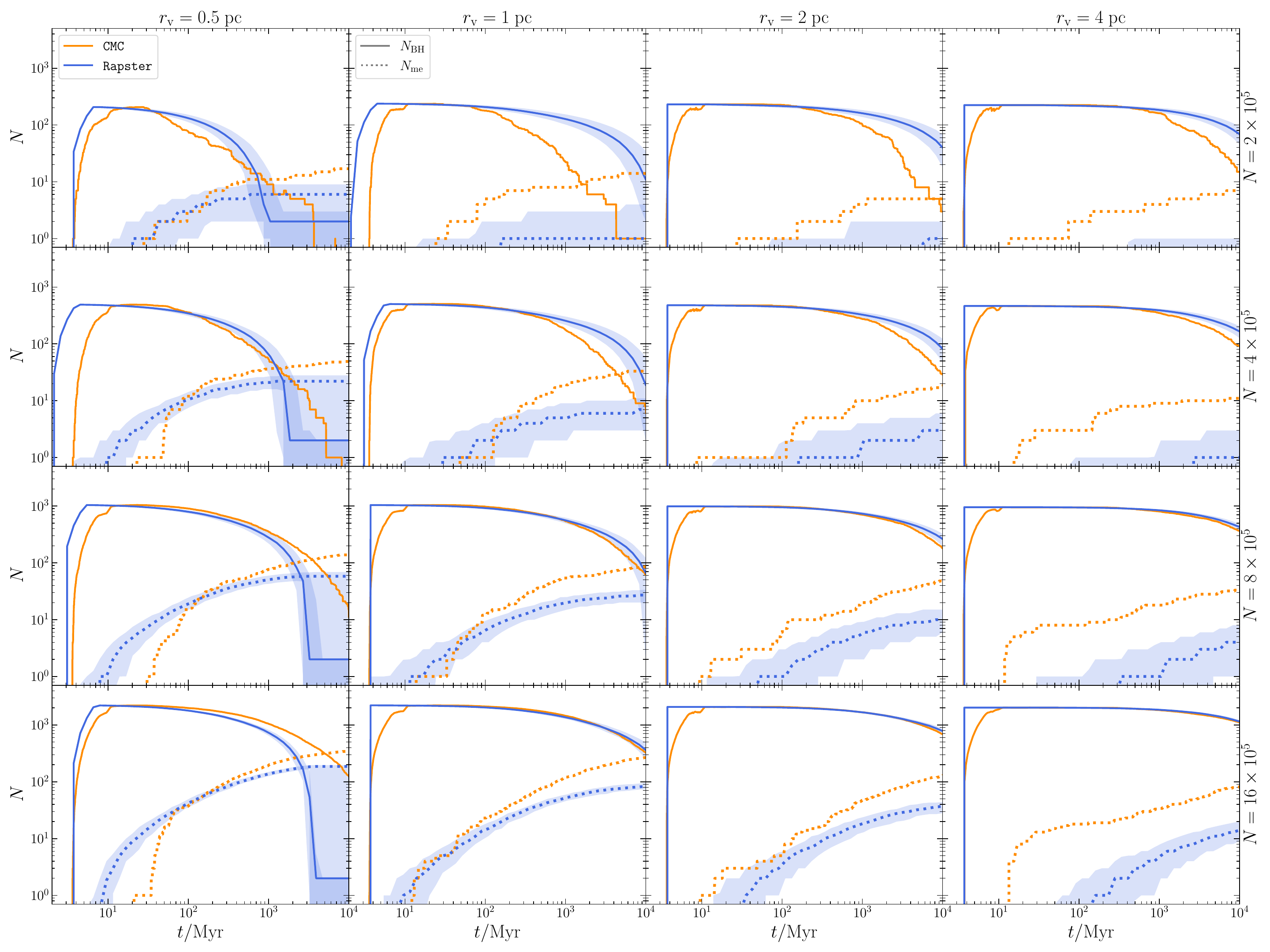}
	\caption{Time evolution of the total number of BHs (solid lines) and of the cumulative number of mergers (dashed lines) on the same grid of models as in Fig.~\ref{Fig:global_evolution}. The bands around the {\tt Rapster} results correspond to the 5, 50, and 95 percentiles.}
	\label{Fig:BH_number}
\end{figure*}

In Fig.~\ref{Fig:BH_number} we plot the total number of BHs (solid lines), as well as the cumulative number of BBH mergers (dotted lines) as a function of time.
For the {\tt Rapster} results we have included ``error bands'' corresponding to the 5, 50, and 95 percentiles estimated on a large set of hundreds of {\tt Rapster} realizations for each model.

\begin{figure*}
	\centering
	\includegraphics[width=1\textwidth]{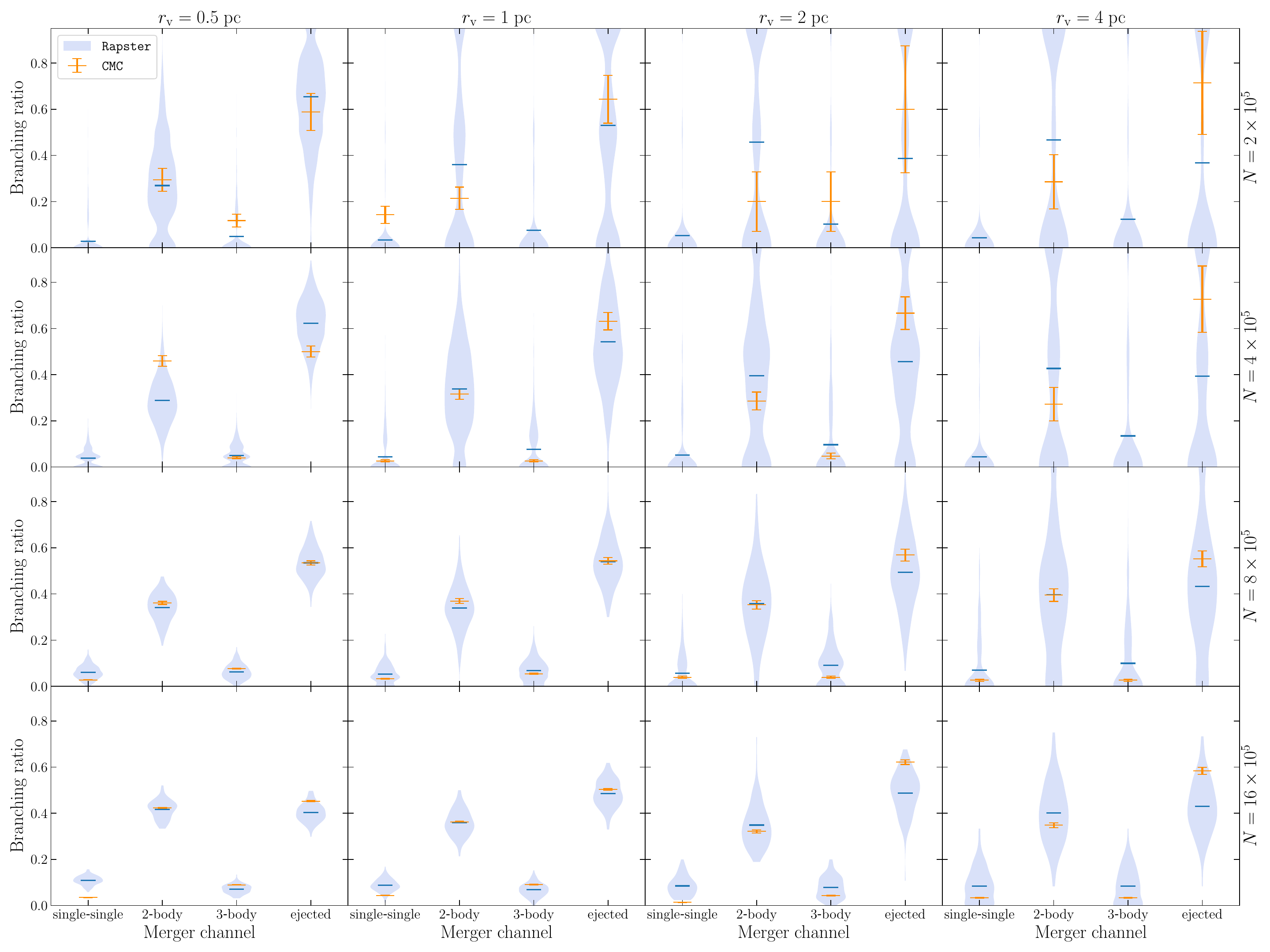}
	\caption{Branching ratios of different merger channels: (a) single-single, (b) 2-body, (c) 3-body, and (d) ejected, for the models of Fig.~\ref{Fig:global_evolution}. For a description of these channels see Sec.~\ref{Sec:Fraction_of_merger_channels}. The horizontal blue lines correspond to the mean fractions of the {\tt Rapster} results, while the violins show the distributions estimated over hundreds of realizations. The orange error bands on the {\tt CMC} fractions have been estimated assuming Poisson statistics. For comparison purposes, 4-body mergers (which correspond to BH--BH captures during BBH--BBH interactions) are discarded.}
	\label{Fig:channels}
\end{figure*}

\begin{figure*}
	\centering
	\includegraphics[width=1\textwidth]{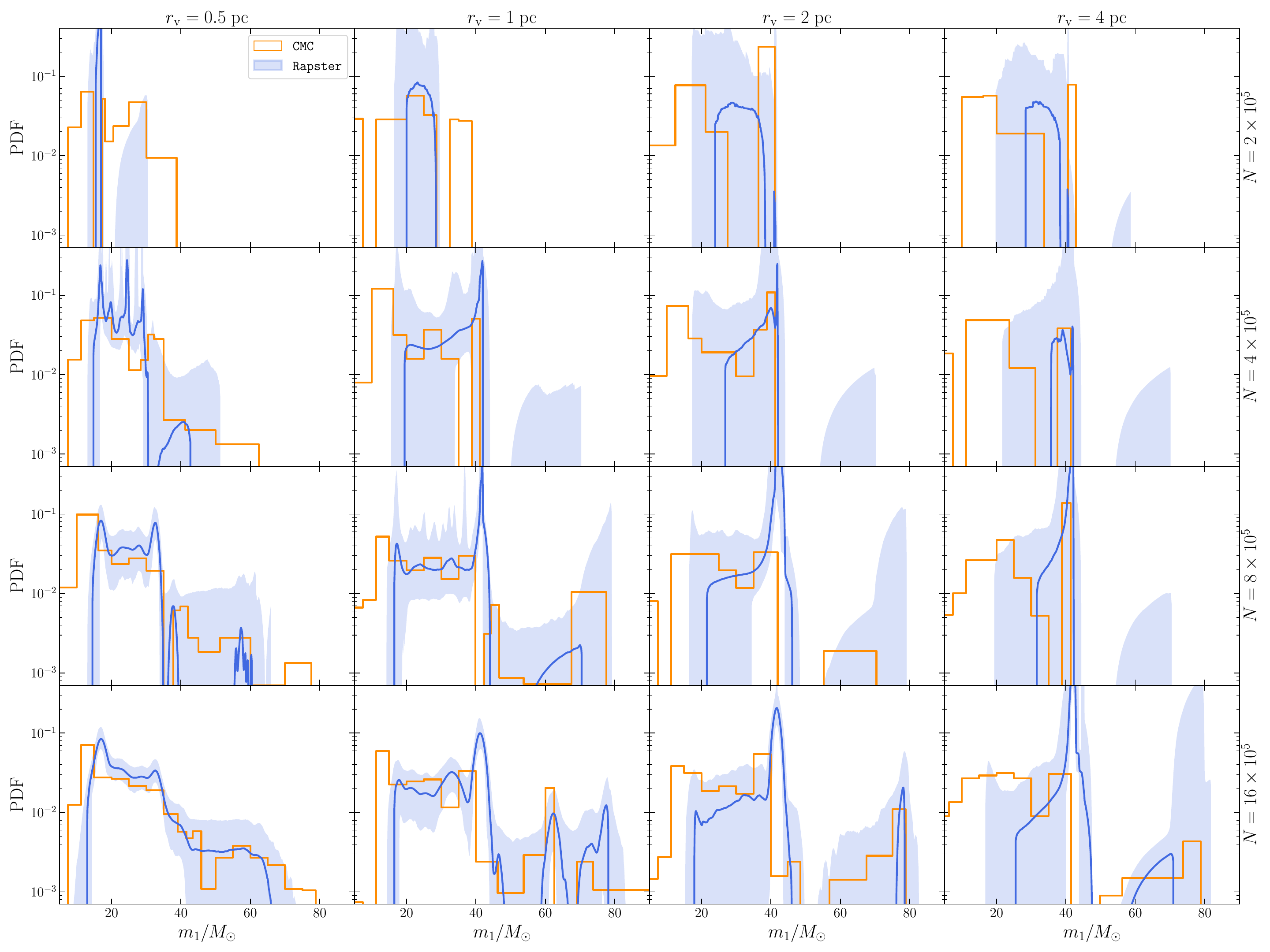}
	\caption{Probability density estimate of the primary mass distribution of BBH mergers for each model in the simulation grid.}
	\label{Fig:primary_mass}
\end{figure*}

\begin{figure*}
	\centering
	\includegraphics[width=1\textwidth]{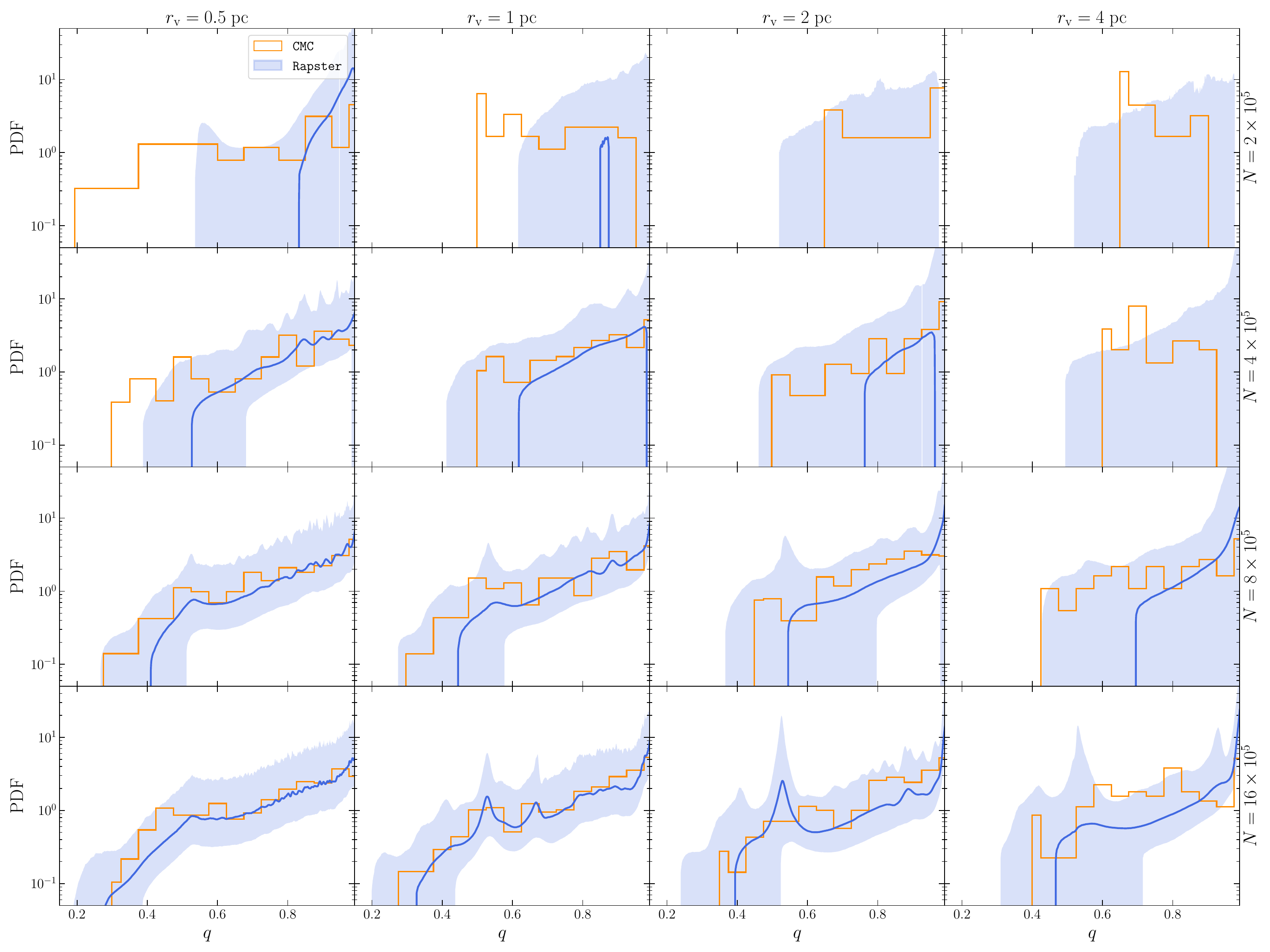}
	\caption{Probability density estimate of the mass ratio distribution of BBH mergers for each model in the simulation grid.}
	\label{Fig:mass_ratio}
\end{figure*}

\begin{figure*}
	\centering
	\includegraphics[width=1\textwidth]{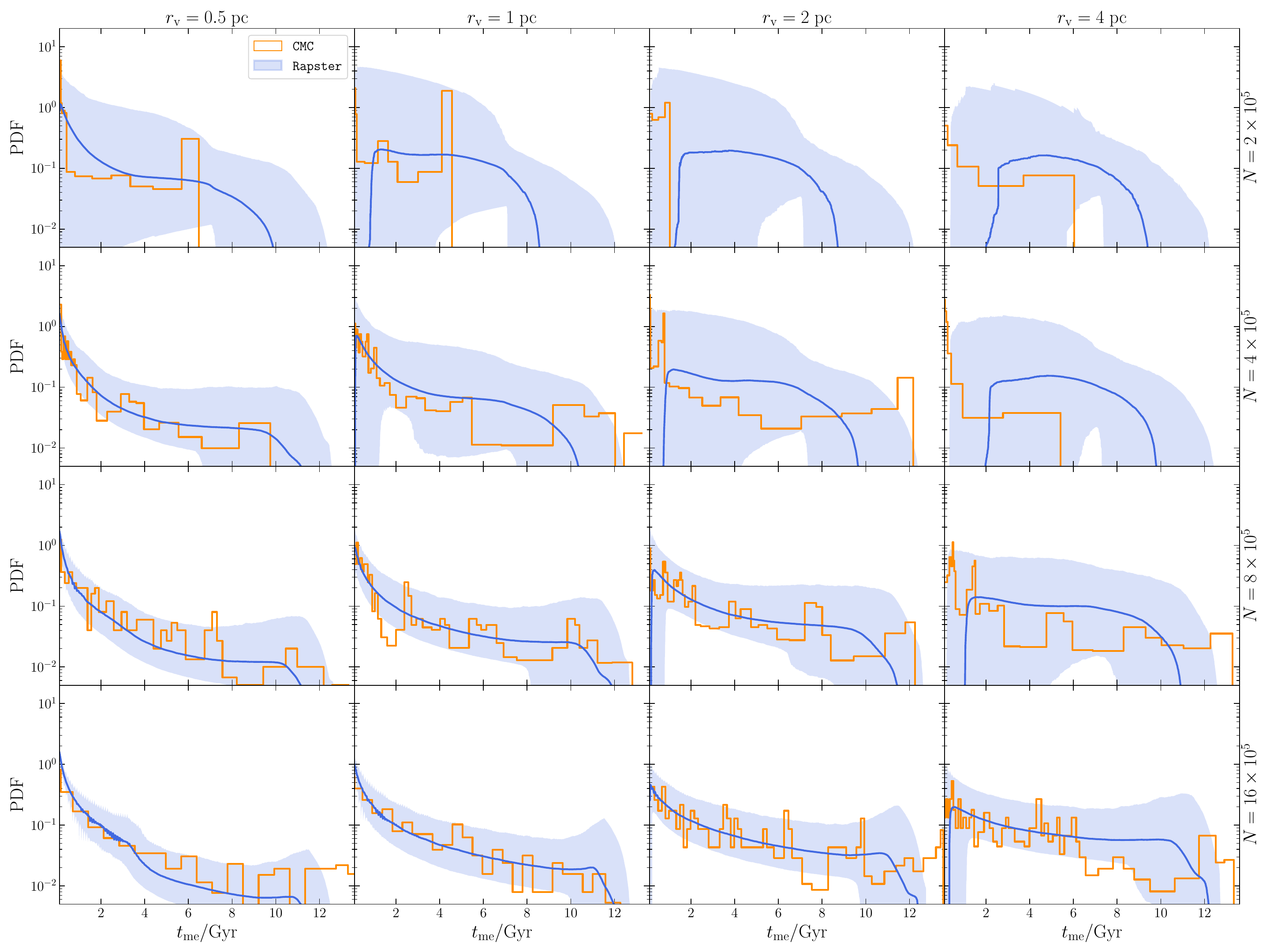}
	\caption{Probability density estimate of the merger time distribution of BBH mergers. Here, merger times are measured from the formation of each star cluster.}
	\label{Fig:Merger_times}
\end{figure*}

\begin{figure*}
	\centering
	\includegraphics[width=1\textwidth]{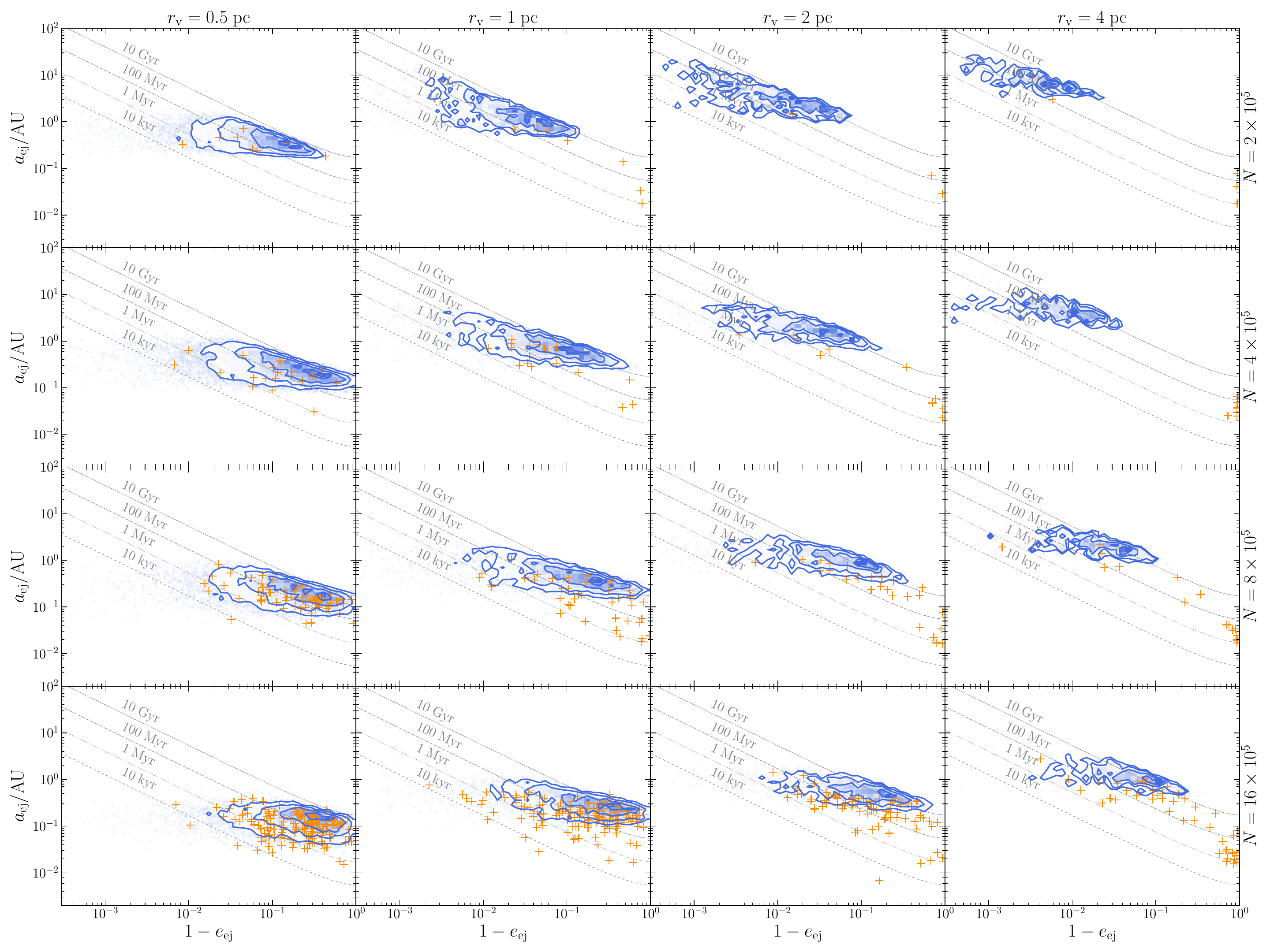}
	\caption{Semimajor axis and eccentricity of BBHs that are ejected and merge outside of the cluster, at the moment of ejection. After ejection, the binary evolves and hardens solely due to GW emission. The dashed gray lines show merger times in the $(1-e_{\rm ej},\,a_{\rm ej})$ parameter space.}
	\label{Fig:ejected_orbits}
\end{figure*}

The more compact systems (those with smaller virial radius $r_{\rm v}$) have smaller relaxation timescales and thus evolve faster. 
As a consequence of the higher central densities, these compact systems form a larger number of BBHs per unit time and have a smaller interaction timescale, which leads to increased BH ejection rates as well as increased merger rates.
In general, all of our $r_{\rm v}=0.5\,\rm pc$ models end up with less than 100 BHs within $10\,\rm Gyr$ (if $N\le16\times10^5$).
Evidently, the BH ejection rate seems to be related with the merger rate. This is no coincidence: the two quantities are density-dependent, and a higher core density leads to a larger interaction rate.
More interactions increase the binary formation rate, and will inevitably lead to a larger number of BH ejections (during BBH--BH encounters).
The evaporation and merger rates progressively decrease as we look at sparser systems with larger values of $r_{\rm v}$.

\subsection{Binary black hole merger properties}
\label{Sec:Binary_black_hole_merger_properties}

In this subsection we describe the properties of BBH mergers in our simulations with an emphasis on the merger channels (\ref{Sec:Fraction_of_merger_channels}), primary mass and mass ratio (\ref{Sec:Primary_mass_and_mass_ratio}), delay times (\ref{Sec:Merger_times}), and orbital properties of ejected merging pairs (\ref{Sec:Orbital_properties_of_ejected_merging_pairs}).

\subsubsection{Fraction of merger channels}
\label{Sec:Fraction_of_merger_channels}

We identify four merger channels:
\begin{enumerate}[label=(\alph*)]
	\item {\it single--single}: corresponds to the GW capture of two single BHs in the core of the BH subsystem;
	\item {\it 2-body}: formation of the BBH through the exchange or 3bb formation channel, and subsequent hardening until the binary decouples from the dynamical environment and merges within the cluster due to GW emission;
	\item {\it 3-body}: GW capture that occurs within the cluster during a strong resonant BBH--BH interaction;
	\item {\it ejected}: ejection of the BBH and merger outside of the cluster within the available time (no larger than the age of the Universe).
\end{enumerate}
This terminology has been borrowed from the {\tt CMC} catalog to facilitate comparison. 
Notice that the {\tt CMC} contains a fifth label for {\it 4-body} mergers, which corresponds to BH--BH captures during BBH--BBH interactions. 
The current version of {\tt Rapster} does not incorporate these merger events, so we discard any 4-body mergers when comparing our results with the {\tt CMC} fractions.
Notice that all but the ``{\it ejected}'' mergers occur in the cluster. Whether the merger remnant is retained or ejected depends on the magnitude of the GW recoil it receives when compared with the cluster escape velocity.

If $N_{\rm me}$ is the total number of mergers in a cluster simulation and $N_{\rm me}^{(i)}$ is the total number of mergers in channel $i$, then the merger fraction of channel $i$ (or ``branching ratio'') is defined as $f_{\rm me}^{(i)}\equiv N_{\rm me}^{(i)}/N_{\rm me}$.
Assuming Poisson statistics for the number of mergers, i.e., that $\delta N_{\rm me}=\sqrt{N_{\rm me}}$ and $\delta N_{\rm me}^{(i)}=\sqrt{N_{\rm me}^{(i)}}$, we write the uncertainty in the merger fractions as $\delta f_{\rm me}^{(i)} = (\sqrt{f_{\rm me}^{(i)}} + f_{\rm me}^{(i)}) / \sqrt{N_{\rm me}}$, which is an upper bound for the error propagation formula with correlated variables.
We use this formula to estimate the uncertainty of the {\tt CMC} merger fractions, that are computed from a single realization for each model. 

In {\tt Rapster}, the exact branching ratio of the merger channels depends on the realization, and thus some scatter is expected due to finite-number statistics.
In Fig.~\ref{Fig:channels} we show the distribution of these fractions as violin plots.
The distribution narrows for heavier clusters because they produce a larger number of mergers within a Hubble time.
This is why the bottom-left panel has a smaller scatter in the violin plots: the heaviest, most compact simulated clusters generate the greatest total number of BBH mergers per simulation.
On the other hand, low-mass clusters may see only a few mergers (or even a single merger) in their evolution: compare the dotted lines in Fig.~\ref{Fig:BH_number}.
This is imprinted in the upper-right panels of Fig.~\ref{Fig:channels}, where the lowest-mass, high-$r_{\rm v}$ clusters usually display a bimodality for the two most popular channels (``2-body'' and ``ejected'').
This is because over a particular realization the cluster may produce a single event (corresponding then to 100\% of the events) in, say, the 2-body channel.
The capture channels (single-single and 3-body) have significantly smaller fractions because they are rare.

Based on the results in Fig.~\ref{Fig:channels}, most of the mergers occur either as 2-body mergers or as ejected mergers, with roughly equal probabilities.
On average, about 20\% of mergers are formed during a single--single or binary--single (3-body) capture episode, again with almost equal probabilities.
In particular, the merger fraction distributions of these capture channels in the $(N,r_{\rm v})=(16\times10^5,0.5\,\rm pc)$ models have negligible support at zero, meaning that those systems are always producing eccentric mergers.
This has interesting consequences for GW astronomy, because the majority of these capture events result in highly eccentric BBH mergers.
Dynamically formed BBHs can be caught early in their inspiral and detected as eccentric binaries in the mHz or deci-Hz band by future or proposed GW observatories. Nonzero eccentricity measurements could potentially help us to distinguish dynamically formed BBHs from those formed through the isolated binary evolution channel~\cite{Breivik:2016ddj, Nishizawa:2016eza, Nishizawa:2016jji, Samsing:2018nxk, Holgado:2020imj, Romero-Shaw:2022xko, Yang:2022tig}.

\subsubsection{Primary mass and mass ratio}
\label{Sec:Primary_mass_and_mass_ratio}

In Figs.~\ref{Fig:primary_mass} and~\ref{Fig:mass_ratio} we show the primary mass and mass ratio distributions, respectively.

The shape of the primary mass distribution depends upon the initial BH mass spectrum and the BH mass sampling probability, that is affected by the cluster dynamics.
Certain features of the initial BH mass spectrum (shown in Fig.~\ref{Fig:BH_spectra}) are imprinted into the mass distribution of mergers: for example, the pileup of BHs at around $40M_\odot$ in the initial mass spectrum is also visible in the primary mass distributions of Fig.~\ref{Fig:primary_mass}.
This particular feature is amplified in the merger component primary mass because there is a dynamical preference for heavier BHs to merge, leading to an overdensity of mergers around that mass scale.
The merger of two $\sim40M_\odot$ BHs produces an $\sim80M_\odot$ BH. If the latter remains in the cluster and merges again, it is likely to become the primary mass of another merger.
Therefore, the existence of a strong peak at $40M_\odot$ in the primary mass distribution is paired with another, relatively weaker peak at $80M_\odot$, which corresponds to mergers involving second-generation BHs.

Dynamical channels and binary--single exchanges lead to a mass-ratio distribution biased toward unity.
Therefore, most BHs that merge dynamically in star clusters are both heavy and of similar mass (Fig.~\ref{Fig:mass_ratio}).
Deviations from this general trend occur when a higher-generation BH merges with a lower-generation one.
For instance, the merger between two first-generation (1g) BHs produces a second-generation (2g)  BH remnant.
If that remnant is retained in the cluster (i.e., if it receives a GW recoil that does not exceed the escape velocity of the cluster) then it may merge again with either a 1g or another 2g BH, thus leading to 1g+2g and 2g+2g events within the cluster.
Since the typical mass of a 2g BH is roughly twice the mass of a 1g BH, the mass ratio distributon of 1g+2g events peaks at $q=0.5$~\cite{Gerosa:2017kvu}.
The small excess in the mass-ratio distribution around $q=0.5$ is caused by these 1g+2g mergers.
The higher-generation BHs tend to be heavier and they contaminate the so-called pair-instability supernova ``upper mass gap,'' which extends from $\sim40M_\odot-120M_\odot$ (although the exact limits are uncertain)~\cite{Franciolini:2024vis}.

\subsubsection{Merger times}
\label{Sec:Merger_times}

In Fig.~\ref{Fig:Merger_times} we plot the merger time distributions.
By merger time, $t_{\rm me}$, here we mean the time of merger relative to the beginning of the simulation.
It is immediately clear that the majority of mergers occur early in the evolution.
This is a consequence of cluster expansion and mass loss, which lead to lower core densities at late times.
Most late-time mergers occur outside of the cluster, after the BHs have been ejected at some earlier time during the cluster's evolution.
Typically, BBHs form early in the evolution of the cluster (while it is still compact). A fraction of them harden efficiently and merge within the cluster, while the rest are ejected and merge on longer timescales in low-density regions due to GW emission.
This picture is more accurate for compact clusters, which expand by a larger factor during their evolution (cf.~Sec.~\ref{Sec:Half_mass_radius}). In less compact clusters the evolution is slower, and a relatively significant fraction of mergers still occur at late times.
These physical processes lead to a variation in the steepness of the merger time distribution which can be seen in the various panels of Fig.~\ref{Fig:Merger_times}. This variation is primarily dependent on $r_{\rm v}$, which plays a more important role than the cluster mass in controlling the relaxation timescale (cf.~Eq.~\eqref{Eq:RelaxationTime}).
In particular, we have verified that models with $r_{\rm v}>0.5\,\rm pc$ have a late-time distribution that follows the $1/t_{\rm me}$ law.

\subsubsection{Orbital properties of ejected merging pairs}
\label{Sec:Orbital_properties_of_ejected_merging_pairs}

Ejected pairs may merge in the low-density field, depending on their orbital parameters. 
In Fig.~\ref{Fig:ejected_orbits} we plot the semimajor axis $a_{\rm ej}$ and the complementary of the eccentricity, $1-e_{\rm ej}$, at the moment when a merging BBH is ejected.
Given those parameters, the subsequent evolution of the binary is uniquely determined (up to an initial phase) by the emission of GWs leading to its merger.

In each panel of Fig.~\ref{Fig:ejected_orbits} we overplot gray dashed contour lines corresponding to constant merger times $\{10\,\rm kyr, 1\,\rm Myr, 100\,\rm Myr, 10\,\rm Gyr\}$.
These contours naturally have a negative slope in the $(1-e_{\rm ej})$--$a_{\rm ej}$ parameter space~\cite{Peters:1963ux,Peters:1964zz}.
This anticorrelation is also observed in the distribution of ejected BBHs: as merging binaries are ejected at a larger semimajor axis, their eccentricity needs to be higher (i.e., $1-e_{\rm ej}$ must be smaller) for them to merge within the available remaining time.
Obviously, there are no points above the $14\,\rm Gyr$ line, because those would correspond to events with a merger time larger than the Hubble time.
The peaks of our predicted distributions, which occur between the ``$100\,\rm Myr$'' and ``$10\,\rm Gyr$'' constant merger time lines, indicate that most ejected pairs merge within a few hundred Myr up to a few billion years following their ejection. These pairs contribute to late-time mergers that originate from the cluster.
Moreover, the observed peak of the distribution shifts to higher values of eccentricity (lower $1-e_{\rm ej}$) as we move toward the left and toward the top in the grid of panels. This is because the escape velocity decreases in this general direction, and binaries tend to be ejected at an earlier stage in their hardening evolution.
Therefore, since the GW merger timescale is a sensitive function of eccentricity~\cite{Peters:1964zz}, only very eccentric binaries can merge within a Hubble time.
As a consequence, the eccentricity distribution of ejected merging pairs is not necessarily thermal.

\section{Conclusions}
\label{sec:conclusions}

In this work we have presented {\tt Rapster}, a new open source {\tt Python} software to carry out rapid simulations of the dynamical assembly of BBHs in dense massive star clusters.
The code takes into account most physical processes relevant to dynamical assembly, including two-, three-, and four-body interactions, and treats them in a semianalytic manner as described in Sec.~\ref{sec:Dynamics}. Simulating a typical globular cluster with {\tt Rapster} takes just a few seconds, and the simulation time scales with the mass of the cluster.

One important limitation of the code is that we neglect original BBHs and BH-star binaries, and their formation through isolated binary evolution. According to observations in the local Universe, the majority of massive stars in young stellar clusters that collapse into BHs tend to form in binaries~\cite{2012ASPC..465..284S,2017ApJS..230...15M}. One way of including original binaries is to use a rapid population synthesis code for field binaries, such as {\tt COMPAS}~\cite{COMPASTeam:2021tbl}, {\tt COSMIC}~\cite{Breivik:2019lmt}, {\tt MOBSE}~\cite{Giacobbo:2017qhh}, or {\tt POSYDON}~\cite{Fragos:2022jik} (among others), to evolve binary stars. While these original binaries are effectively a separate channel and could boost the merger rates, we have decided to focus on the dynamical formation of binaries in this version of the code, and compute the single BH mass spectrum instead.

In simulating star clusters, there is a trade-off between computational performance and accuracy. 
The method we use to simulate the BH subsystem in star clusters relies on the theory developed by Breen and Heggie (2013)~\cite{Breen:2013vla}, which breaks down when the cluster contains a relatively small number of BHs (roughly less than 40, although the exact number depends on many parameters). Furthermore, near the complete evaporation of the BH subsystem and as $N_{\rm BH}\to0$, the Coulomb logarithm is known to diverge, so our approximation of fixing $\log\Lambda_{\rm BH}=1$ may not be accurate. This limit is also reached in low-mass clusters ($M_{\rm cl}\lesssim10^5M_\odot$), where the initial number of BHs is only a few tens. Therefore, the prescriptions used in {\tt Rapster} may become unreliable for light clusters.

A more accurate modeling of dense stellar environments requires $N$-body codes that implement sophisticated direct integration techniques or the solution of Fokker-Planck equations with Monte Carlo methods, but these approaches require considerable computational power. 
Furthermore, the exact initial conditions of star clusters are uncertain. Due to the chaotic character of these systems, even very accurate $N$-body simulations may yield different results, depending on whether residual gas dynamics or other physical effects are taken into account.
{\tt Rapster} comes in handy when there is a necessity to quickly obtain a large catalog of results with reasonable accuracy. An example of its application in the future is the inference of BBH population hyperparameters, given large catalogs of actual BBH merger events from GW observatories.
The other benefit of a semianalytic code such as {\tt Rapster} is that one can efficiently explore the dependence of the results on the parameter space of initial conditions. By simulating a very large sample of star clusters one can also probe better the tails of the BBH merger distributions, as well as other ``minor'' features that depend on the dynamics.

We have performed a rather extensive comparison of our results with the {\tt CMC} code over a grid of star cluster models.
The evolution of the cluster and general trends with particle number $N$ and virial radius $r_{\rm v}$ were verified and explained based on the physical ingredients included in our model.
While our results are in generally reasonable agreement with {\tt CMC}, we note two important differences.

The first difference is that our cumulative number of mergers is consistently smaller than the one predicted by {\tt CMC}, on average by a factor of a few, with the largest deviations cooresponding to low-mass, loose clusters. While a number of mergers in {\tt CMC} that we do not see in {\tt Rapster} come from the isolated binary evolution channel and contribute in the low-mass regime, the omission of this channel in {\tt Rapster} does not account the entirety of the observed discrepancy in the rates.
Despite this variation, the predicted merger rate density can vary by almost two orders of magnitude (see Fig.~3 in Ref.~\cite{Mandel:2021smh}, under the panel ``Globular clusters''), with a significant contribution to the uncertainty coming from the (unknown) number density of globular clusters across cosmic time.

The second main difference is that our primary BBH mass distribution function peaks at the pile-up ($\sim40M_\odot$), and it does not display a strong low-mass peak at $\sim20M_\odot$. Again, this feature could be partly caused by the absence of mergers resulting from primordial binary stars, because a good fraction of these mergers have low mass. 
Moreover, recent $N$-body results from the {\tt DRAGON-II} simulations (Fig.~3 in Ref.~\cite{Sedda:2023qlx}) demonstrate that most dynamically formed BBH mergers occur around $m_1\sim m_2\sim 40M_\odot$, while isolated binary evolution is responsible for low-mass mergers -- with the caveat that those simulations were carried out for at most $\sim2\,\rm Gyr$, without reference to late-time mergers. 
In addition, our results are consistent with Ref.~\cite{Antonini:2022vib}, which predicts a similar distribution where most dynamical mergers occur near the pulsational pair-instability pile-up.

Throughout this work we have assumed a minimum binary hardness of $\eta_{\rm min}=5$.
While this is a conservative choice (see~\cite{Morscher:2014doa}), it is important to point out that the 3bb formation, BH ejection, and BBH merger rates all depend on $\eta_{\rm min}$.
Specifically, a larger value of $\eta_{\rm min}$ in the model means that binaries require a higher 3bb rate to form, since the critical encounter pericenter is inversely proportional to $\eta_{\rm min}$.
To achieve this, the balanced evolution condition requires a much denser BH core, which inevitably boosts the interaction rate -- and hence, the merger and BH ejection rates (see the last paragraph of Appendix~\ref{app:ThreeBodyBinaryRate} for a more formal discussion).
Therefore, as we increase $\eta_{\rm min}$, the lifetime of the BH subsystem is shortened.
Most certainly, in reality binaries form with all possible values of  $\eta>0$. For this reason, any choice of $\eta_{\rm min}$ is artificial and leads to unphysically large densities.
However, soft binaries with $\eta<1$ do not contribute to cluster heating because they tend to soften according to the Heggie-Hills law, while those with $\eta$ marginally close to 1 are vulnarable to ionization and softening.
Therefore here we have not tried to estimate the best value of $\eta_{\rm min}$. Instead we have made a conservative choice that does not assume an unreasonably large value of $\eta_{\rm min}$, which could lead to unreasonably high BH core densities.

In this paper we were only concerned with the {\em intrinsic} properties of astrophysical BBH mergers (mass, spin distributions, and coalescence rate). The question of detectability and measurability of dynamical BBHs by current and future GW detectors is left for future work.
Studying the dependence of the BBH population on different remnant-mass prescriptions requires a careful analysis that goes beyond the scope of this work. This will be the subject of a future study.
Different remnant-mass prescriptions affect the initial BH mass spectrum, and therefore also the mass distribution of BBH mergers.
We also plan to carry out a more thorough comparison between {\tt Rapster} and the other semianalytic simulation codes mentioned in the Introduction: {\tt B-POP}, {\tt FASTCLUSTER} and {\tt cBHB}.
This is an important task, because it is possible (and likely) that different semianalytic codes will lead to different predictions. The comparison with real data is therefore likely to be limited by the different physical assumptions made in each code.
\\

The supporting data for this paper are openly available in~\url{https://zenodo.org/records/10626210}.

\acknowledgments
We are specially grateful to Kyle Kremer and Newlin Weatherford for help in understanding and interpreting the {\tt CMC} data.
We also thank Carl Rodriguez, Mark Cheung, Ken Ng, Nathan Steinle, Andrea Antonelli, Giacomo Fragione, Roberto Cotesta, Luca Reali, Veome Kapil, Gabriele Franciolini, Nicholas Stone, Floor Broekgaarden, Dany Atallah, Miguel Martinez, Frederic Rasio, Mike Zevin, and Fabio Antonini for useful discussions and comments.
This paper is supported by the Onassis Foundation - Scholarship ID: F ZT 041-1/2023-2024.
K.K., V.S. and E.B. are supported by NSF Grants No. AST-2006538,  No. PHY-2207502, No. PHY-090003, and No. PHY-20043, by NASA Grants No. 20-LPS20-0011 and No. 21-ATP21-0010, by the John Templeton Foundation Grant 62840, by the Simons Foundation, and by the Italian Ministry of Foreign Affairs and International Cooperation Grant No.~PGR01167.
This work was carried out at the Advanced Research Computing at Hopkins (ARCH) core facility (\url{rockfish.jhu.edu}), which is supported by the NSF Grant No.~OAC-1920103.
The authors acknowledge the Texas Advanced Computing Center (TACC) at The University of Texas at Austin for providing {HPC, visualization, database, or grid} resources that have contributed to the research results reported within this paper~\cite{10.1145/3311790.3396656}. URL: \url{http://www.tacc.utexas.edu}

\appendix

\section{BH core radius}
\label{app:BHcoreRadius}

We derive the core radius of the BH subsystem necessary for a balanced evolution from first principles, based on the methodology of Breen and Heggie~\cite{Breen:2013vla} (see also Chap. 23 of Ref.~\cite{2003gmbp.book.....H}, pages 243-244).
Balanced evolution is expressed by the condition that the energy generated in the core must be balanced by the heat flow rate in the system (in our case, in the BH subsystem):
\begin{align}
	\Gamma_{\rm 3bb}Q_{\rm BBH}={\zeta_{\rm BH}|E_{\rm BH}|\over\tau_{\rm rh,BH}}.
	\label{Eq.Balanced_evolution}
\end{align}
Here we have written the energy generation rate as the product of the 3bb formation rate $\Gamma_{\rm 3bb}$ (the rate at which hard BBHs form, derived in Appendix~\ref{app:ThreeBodyBinaryRate} below) and $Q_{\rm BBH}$, the energy generated per hard BBH via hardening in the core. The total energy of the BH subsystem is $|E_{\rm BH}|=0.2G(N_{\rm BH}\overline{m}_{\rm BH})^2/r_{\rm h,BH}$, by the virial theorem.

During a binary--single interaction, a fraction of the binary's binding energy, which is proportional to the binding energy [see Eq.~\eqref{Eq:Binding_energy_change}], is transformed into external translational degrees of freedom (kinetic energy). After the encounter, both the binary and the single recoil.
By conservation of momentum, the single receives the largest velocity kick and is more prone to ejection from the cluster.
As the binary hardens, the energy released during the encounters increases proportionally. At some point, when the semimajor axis becomes smaller than a critical value $a_{\rm cr}$, the binary becomes an ``ejector,'' meaning that each successive interaction results in the ejection of the single.
Consequently, the binary hardens until it reaches another critical radius, $a_{\rm ej}$, at which point the energy released is so high that the binary is able to eject itself, i.e., it becomes a ``self-ejector'' (using the terminology of Ref.~\cite{2010ARA&A..48..431P}).
As long as $a>a_{\rm cr}$, the recoiling single is not ejected, and the excess energy it carries can be shared among the members of the system within one relaxation time, i.e., the binary heats up the system with an efficiency of 100\%. 
While $a_{\rm ej}<a<a_{\rm cr}$, only about $\sim30\%$ of the energy released is converted into kinetic energy of the binary and it is conducted throughout the system through relaxation.
The other $\sim70\%$ of the energy is locked into the ejected single BH and cannot be conducted among other members of the system, because the single escapes in less than a crossing time from the cluster.
In this order-of-magnitude estimate, we have assumed that all BH particles are of the same mass $\overline{m}_{\rm BH}$.
In conclusion, by conservation of energy, the amount of energy generated by a single BBH available to be shared among all the BHs in the subsystem is
\begin{align}
	Q_{\rm BBH} &\simeq {G\overline{m}_{\rm BH}^2\over2a_{\rm cr}} + {1\over3}\left( {G\overline{m}_{\rm BH}^2\over2a_{\rm ej}} - {G\overline{m}_{\rm BH}^2\over2a_{\rm cr}} \right) \nonumber \\ &={G\overline{m}_{\rm BH}^2\over6}\left( {2\over a_{\rm cr}} + {1\over a_{\rm ej}} \right),
	\label{Eq.Heat_generated_1}
\end{align}
where we have neglected the energy of the binary when it formed near the hard-soft boundary. In the following we estimate both $a_{\rm cr}$ and $a_{\rm ej}$.

For simplicity, let us consider a binary--single interaction where all particles have the same average BH mass, $\overline{m}_{\rm BH}$.
By momentum conservation, the single will have a velocity $v_{S}'$ that is double the recoil of the binary $v_{B}'$ after the encounter: $v_{S}'=2v_{B}'$ (assuming for simplicity that the binary and the single exit in opposite directions). 
By energy conservation, $(1/2)\mu v_{\rm rel}^2 - |E_{\rm BBH}|=(1/2)\mu v_{\rm rel}'^2 - |E_{\rm BBH}'|$, where $\mu=2\overline{m}_{\rm BH}/3$ is the reduced mass of the system and $v_{\rm rel}$ ($v_{\rm rel}'$) is the relative velocity before (after) the interaction, respectively.
Moreover, $|E_{\rm BBH}'|-|E_{\rm BBH}|=\beta G\overline{m}_{\rm BH}^2/(4a)$ is the binding energy change [see Eq.~\eqref{Eq:Binding_energy_change}].
The relative velocity before the interaction is $v_{\rm rel}^2 = v_{B}^2 + v_{S}^2 = (3/2)\langle v_{\rm BH}^2 \rangle$, while $v_{\rm rel}'=v_{B}'+v_{S}'=3v_{B}'=(3/2)v_{S}'$.
We have also defined $\xi=\overline{m}_{\rm BH}\langle v_{\rm BH}^2\rangle / (\overline{m}\langle v_\star^2 \rangle)$ to be the temperature ratio (see Sec.~\ref{sec:SegragationAndVelocities}) and $v_{\rm esc}=2\langle v_\star^2 \rangle^{1/2}$.
Thus $v_{\rm esc}^2 = 4p\langle v_{\rm BH}^2\rangle/\xi$, where $p=\overline{m}_{\rm BH}/\overline{m}$.
To solve for $a=a_{\rm cr}$ or $a=a_{\rm ej}$, we set $v_{\rm S}'=v_{\rm esc}$ or $v_{\rm B}'=v_{\rm esc}$, respectively.

After some algebra we obtain:
\begin{subequations}
	\begin{align}
		a_{\rm cr}&={\beta G\overline{m}_{\rm BH}\over 2\langle v_{\rm BH}^2\rangle} \left({6p\over\xi} - 1\right)^{-1},  \\
		a_{\rm ej}&={\beta G\overline{m}_{\rm BH}\over 2\langle v_{\rm BH}^2\rangle} \left({24p\over\xi} - 1\right)^{-1}.
	\end{align}
\end{subequations}
Plugging these expressions for $a_{\rm cr}$ and $a_{\rm ej}$ into Eq.~\eqref{Eq.Heat_generated_1} above yields
\begin{align}
	Q_{\rm BBH}={1\over\beta}\left( {12p\over\xi} - 1 \right)\overline{m}_{\rm BH}\langle v_{\rm BH}^2\rangle.
\end{align}
By plugging in some typical values of $\beta=4/7$, $p\sim30$, and $\xi\sim4$, we get $Q_{\rm BBH}\sim 150 \langle m_{\rm BH}v_{\rm BH}^2\rangle$. 
This is larger than the value quoted in~\cite{2003gmbp.book.....H} (page 244) by a factor of $\sim4$.
This is expected, because in our scenario BBHs harden in the core of the BH subsystem, which is embedded within the deeper potential of the stars.
As such, the escape velocity in this two-component system is larger from the point of view of the source gravity of the BH subsystem.
Therefore, BBHs can harden to even larger values of hardness, releasing a greater amount of energy in the system.

Finally, combining everything into Eq.~\eqref{Eq.Balanced_evolution} and assuming that the BH system follows the isothermal distribution~\cite{Breen:2013vla}, we get
\begin{align}
	{r_{\rm c,BH}\over r_{\rm h,BH}} &= N_{\rm BH}^{-2/3} \left( { {\cal C} \over \zeta\ln\Lambda_{\rm BH} } \right)^{1/3},
\end{align}
where
\begin{align}
	{\cal C} = {1.76\over\beta\psi_{\rm BH}} \left( {12p\over\xi} - 1 \right)\eta_{\rm min}^{-{11\over2}}(1+3\eta_{\rm min})(1+6\eta_{\rm min}).
\end{align}

Notice that Eq.~\eqref{Eq.Balanced_evolution} is only an approximation for various reasons. First of all, it assumes that hard binaries come only from 3bb formation. While it is true that most BBHs form in the core through three-body encounters, a fraction of them assemble via the exchange channel (star-star$\to$BH-star$\to$BH-BH): BBHs that form from a two-body capture merge rapidly, and do not have enough time to contribute to heating the system.
Moreover, the balanced condition we consider assumes that all formed binaries harden all the way to $a_{\rm ej}$, while some of them become ionized way before reaching that point, thus leading to a slight overestimation of the energy generated in the core.
Last but not least, this is only an averaged description of the process, because we have used the mean mass and velocity dispersion of BHs.

\section{Three-body binary rate}
\label{app:ThreeBodyBinaryRate}

In this appendix we derive the 3bb rate, which is crucial for calculating the 3bb formation timescale in the core of the BH subsystem (see Sec.~\ref{sec:ThreeBodyBinaryFormation}).

The two-body interaction rate density between masses $m_1$ and $m_2$, with relative velocity at infinity $v_{1,2}$ and pericenter distance $r_{p_{1,2}}$, is
\begin{align}
	\gamma_{1,2} = n_1n_2\left\langle v_{1,2}\pi r_{p_{1,2}}^2\left(1 + {2Gm_{12}\over r_{p_{1,2}}v_{1,2}^2}\right) \right\rangle,
\end{align}
where the angular brackets represent a velocity average over the Maxwellian distribution with one-dimensional velocity dispersion $\sigma_{1,2}$, $m_{12}=m_1+m_2$, and $n_1$ and $n_2$ are the number densities of the two interacting species, respectively.
Using the fact that $\langle v_{1,2}\rangle=\sigma_{1,2}\sqrt{8/\pi}$ and $\langle v_{1,2}^{-1}\rangle=\sigma_{1,2}^{-1}\sqrt{2/\pi}$, we obtain
\begin{align}
	\gamma_{1,2} = 2\sqrt{2\pi} n_1n_2 r_{p_{1,2}}^2 \sigma_{1,2}\left( 1 + {Gm_{12}\over r_{p_{1,2}}\sigma_{1,2}^2} \right).
\end{align}
Similarly, the interaction rate of a third mass $m_3$ with $m_{12}$ with relative one-dimensional velocity dispersion $\sigma_{12,3}$ at infinity and pericenter distance $r_{p_{12,3}}$ is
\begin{align}
	\Gamma_{12,3} = 2\sqrt{2\pi} n_3 r_{p_{12,3}}^2\sigma_{12,3}\left( 1 + {Gm_{123}\over r_{p_{12,3}}\sigma_{12,3}^2} \right),
\end{align}
where $n_3$ is the number density of objects with mass $m_3$, and $m_{123}=m_{12}+m_3$. Moreover, $\sigma_{12,3}^2 = \sigma_{12}^2+\sigma_3^2$. Thus, using equipartition and assuming that all particles have the same mass and velocity dispersion, we have $\sigma_{12}^2=\sigma_1^2/2$, and hence $\sigma_{12,3}^2=(3/2)\sigma_1^2$.

The rate density of three single particles $m_1$, $m_2$, and $m_3$ closely interacting simultaneously is given by the product of the two-body rate density $\gamma_{1,2}$ times the probability that the third particle finds itself in the vicinity of the interacting 1--2 pair, summed over all dynamically independent permutations: $\gamma_{1,2,3}=(\gamma_{1,2}p_{3}+\gamma_{2,3}p_1+\gamma_{3,1}p_2)/3$, where we have also included a normalization factor of 3.
The latter probability is conditional, and it is given by the product of the two-body rate $\Gamma_{ij,k}$ with the passage timescale of i--j: $p_k=\Gamma_{ij,k}\tau_{i,j}$.
The symbol $\tau_{i,j}$ is given by $2r_{p_{i,j}}/v_{p_{i,j}}$, where $v_{p_{i,j}}=\sqrt{2Gm_{ij}/r_{p_{i,j}}}$ is the relative velocity at the point of closest approach (or relative pericenter velocity).
Putting everything together yields:
\begin{align}
	\gamma_{1,2,3} &= {8\pi\over3} n_1n_2n_3 r_{p_{1,2}}^2 r_{p_{12,3}}^2 \sigma_{1,2}\sigma_{12,3} \sqrt{r_{p_{1,2}}^3\over Gm_{12}} \nonumber \\ &\times \left(  1 + {Gm_{12}\over r_{p_{1,2}}\sigma_{1,2}^2}  \right) \left( 1 + {Gm_{123}\over r_{p_{12,3}}\sigma_{12,3}^2}   \right) \nonumber \\ &+ {\rm cycles\ of\ [1,2,3]}.
	\label{eq:3b_rate_formula}
\end{align}

We make some further simplifying assumptions.
In Eq.~\eqref{eq:3b_rate_formula} we set all masses equal to the mean BH mass $\overline{m}_{\rm BH}$, we set all one-dimensional relative velocity dispersions to be $\sqrt{2/3}\langle v_{\rm BH}^2\rangle^{1/2}$, and we also set $n_1=n_2=n_3=n_{\rm c,BH}$, the core BH number density.
Moreover, we assume that the triple encounter takes place in a region of size $\sim a$, thus we set all pericenter distances to $a$.
Lastly, we define the minimum hardness parameter to be $\eta_{\rm min}\equiv G\overline{m}_{\rm BH}/(a\langle v_{\rm BH}^2\rangle)$.
Taking all of these assumptions and definitions into consideration, we write the three-body rate density as
\begin{align}
	\gamma_{1,2,3}(\eta\ge\eta_{\rm min}) &= {8\pi\over\sqrt{3}} \eta_{\rm min}^{-11/2}(1+3\eta_{\rm min})\left(1 + 6\eta_{\rm min}\right) \nonumber \\ &\times n_{\rm c,BH }^3 {(G\overline{m}_{\rm BH})^5\over \langle v_{\rm BH}^2\rangle^{9/2}},
\end{align}
where the condition ``$\eta\ge\eta_{\rm min}$'' means that the orbits of the three interacting BHs are all focused in a region with a size of at most $G\overline{m}_{\rm BH}/(\eta_{\rm min}\langle v_{\rm BH}\rangle)$.
To get the total rate we integrate the rate density over the core volume, or (since $n_{\rm c,BH}$ is constant over the core) we simply multiply by the volume of the core: 
\begin{align}
	\Gamma_{1,2,3}(\eta\ge\eta_{\rm min})={4\pi\over3}r_{\rm c,BH}^3\gamma_{1,2,3}(\eta\ge\eta_{\rm min}), 
	\label{Eq.Total_3bb_rate}
\end{align}
where $r_{\rm c,BH}$ is the BH core radius (cf.~Eq.~\eqref{Eq.BHcoreRadius}).
Since we are interested in binary formation with hardness greater than unity (leading to formation of hard binaries), we assume that all three-body interactions are energetic enough to lead to the formation of a binary. Therefore, $\Gamma_{\rm 3bb}=\Gamma_{1,2,3}$.
Finally, the 3bb timescale is given by $\tau_{\rm 3bb}=\Gamma_{\rm 3bb}^{-1}$.

Our formalism for the 3bb rate is similar to Refs.~\cite{Morscher:2014doa,Rodriguez:2021qhl}.
In particular, our result agrees very well with Eq.~(2) of Ref.~\cite{Morscher:2014doa}, with two main differences: (i) we find a factor of $(1+3\eta_{\rm min})(1+6\eta_{\rm min})$ instead of $(1+2\eta_{\rm min})(1+3\eta_{\rm min})$, and (ii) the overall normalization is $8\pi/\sqrt{3}$ instead of $2\pi^2$, i.e., it is off by a factor of $\simeq1.4$.
Nevertheless, these differences collectively result in an overall variation in the 3bb rate on the order of no more than $40\%$ for $\eta_{\rm min}\ge2$.
We have checked that these differences do not significantly affect the evolution of star clusters.

Finally, we calculate the net dependence of the 3bb rate on $\eta_{\rm min}>1$.
We have shown above that $\Gamma_{\rm 3bb}$ depends directly on a decreasing function of $\eta_{\rm min}$, $f(\eta_{\rm min})\equiv\eta_{\rm min}^{-11/2}(1+3\eta_{\rm min})(1+6\eta_{\rm min})$.
Thus one might naively conclude that as $\eta_{\rm min}$ is increased, the binary formation rate becomes inefficient, since the critical pericenter for binary formation has effectively decreased.
This interpretation, however, is only correct if the number density is kept fixed.
In the framework of balanced evolution, which is the main assumption of {\tt Rapster}, a lower binary formation rate should have a negative feedback in the BH core radius $r_{\rm c,BH}$.
The latter should decrease so that close triple BH interactions become frequent enough to generate the necessary energy for the cluster to support itself and expand (cf.~Eq.~\eqref{Eq.Balanced_evolution}).
As we showed in Appendix~\ref{app:BHcoreRadius}, the cube of the BH core radius depends directly on the same function $f(\eta_{\rm min})$.
Therefore, since $\Gamma_{\rm 3bb}\propto f(\eta_{\rm min}) n_{\rm c, BH}^3$ and $n_{\rm c,BH}\propto r_{\rm c,BH}^{-3}$, we finally have $\Gamma_{\rm 3bb}\propto f(\eta_{\rm min})^{-2}$.
This means that, since $f(\eta_{\rm min})$ is a monotonically increasing function for $\eta_{\rm min}>1$, the 3bb rate effectively increases as the fictitious $\eta_{\rm min}$ parameters grows, because it has a strong dependence on the BH core density.

\section{Contribution of BH-BH-star three-body encounters to BBH formation}
\label{app:3bbWithStarAgent}

The three-body rate formula $\Gamma_{3b}$ calculated in Eq.~\eqref{Eq.Total_3bb_rate} as a function of the masses involved in the three-body interaction gives the rate for three bodies to encounter in a predetermined region, and not the binary formation rate itself.
However, as shown in Ref.~\cite{1976A&A....53..259A}, the probability of binary formation during a three-body encounter of equal masses asymptotically approaches $100\%$ when the region of interaction is small enough.
In our case, as already mentioned in Sec.~\ref{sec:ThreeBodyBinaryFormation}, we consider binaries forming with a hardness of at least $\eta_{\rm min}=5$, and therefore triple encounters of BHs would almost always result in the formation of a BBH.
It would also seem reasonable to include the triple interaction of two BHs and a single star, which mostly plays the role of a catalyst for BBH assembly.
Nevertheless, binary assembly through triple interactions of the kind BH--BH--star is rare, as discussed below.

In general, the rate of binary formation via this channel would be given by $\Gamma_{\rm 3bb}={\cal Q}(m_1,m_2,m_3,\eta)\times\Gamma_{3b}$\,, where ${\cal Q}(m_1,m_2,m_3,\eta)\in[0,1]$ is an efficiency factor that depends on mass and hardness, and characterizes the fraction of three-body encounters that induce a bounded pair~\cite{Kritos:2020wcl}. 
To the best of our knowledge, this factor has not been calculated before. Here we present only a crude estimate.

One way to treat a three-body encounter of bodies A, B and C is to split it in two nested interactions: a single--single, nearly parabolic interaction between A and B, and a binary--single encounter between the single C and the metastable pair AB with positive energy given by the relative kinetic energy $T_{\rm AB}=0.5\mu_{\rm AB}\boldsymbol{v}_{\rm A,B}^2$, where $\mu_{\rm AB}$ is the reduced mass, and $\boldsymbol{v}_{\rm A,B}$ the relative velocity of A and B.
Body C in the outer interaction serves as an ``external agent'' whose role is to extract enough energy for the pair AB to become a bound binary system.

The AB pair is not bound, however its energy at pericenter is (by conservation of energy) equal to its energy at infinity, $T_{\rm AB}$.
Therefore, if we approximate the AB--C encounter as a binary--single interaction, then the average amount of energy lost by the (yet unbound) AB pair during the outer binary--single encounter is $|\langle\Delta E\rangle|/T_{\rm AB}$, where the numerator is given by Eq.~\eqref{Eq.HardeningEq}.
Using the definition of binary hardness [cf.~Eq.~\eqref{Eq:hardenessRatio}] we conclude that $|\langle\Delta E\rangle|/T_{\rm AB}\sim\eta_{\rm AB}\beta m_{\rm C}/m_{\rm AB}$, where $m_{\rm AB}=m_{\rm A}+m_{\rm B}$ is the mass of the candidate binary.
The only mass dependence of this fractional energy change is through the ratio $m_{\rm C}/m_{\rm AB}$~\cite{1980AJ.....85.1281H}.
If $|\langle\Delta E\rangle|/T_{\rm AB}>1$, then the pair AB becomes a bound system, otherwise A, B, and C fly away as singles.
Hence, in this crude approximation, a bound pair with the minimum hardness can only be induced if $\eta_{\rm min}\beta m_{\rm C}>m_{\rm AB}$.
This leads to:
\begin{align}
{\cal Q}(m_{\rm A},m_{\rm B},m_{\rm C},\eta)\sim
\begin{cases}
&1,{\ \rm if\ } \eta_{\rm min}\beta m_{\rm C}>m_{\rm AB},\\
&0,{\ \rm otherwise}.
\end{cases}
\label{Eq.Proba3bb}
\end{align}
Note that, even though the estimate implies an abruptly varying factor, this probability is expected to vary smoothly as a sigmoidlike function over the domain, asymptotically approaching the values given in Eq.~\eqref{Eq.Proba3bb}.

As an example, if $\eta\gg1$, then we can approximate the hardening rate constant as $\beta=4/7$~\cite{Samsing:2017xmd}.
For two BHs of equal mass $m_{\rm A}=m_{\rm B}=10M_\odot$, a BBH is likely to form only if the mass of the ``external agent'' is greater than $m_{\rm C}^{\rm min} = 20M_\odot/(\eta_{\rm min}\beta)\approx 7 M_\odot$, taking $\eta_{\rm min}=5$.
Thus, the encounter between two BHs (inner interaction) and a single third star with a mass of 1$M_\odot$ (outer interaction) is unlikely to produce a BBH.
The situation is even worse for heavier BHs, and the only way to produce a bound pair at a non-negligible rate is via a three-BH encounter. In conclusion, we can neglect the formation of BBHs via triple BH--BH--star interactions in our code.

\section{Gravitational captures}
\label{app:GravitationalCaptures}

The average timescale for two-body captures is given by Eq.~\eqref{Eq.capTime}.
A pair of initially unbound BHs with masses $m_1$ and $m_2$ may form a bound system because of relativistic dissipative effects if the impact parameter $b$ of the encounter does not exceed a maximum value given by Eq.~(17) of Ref.~\cite{OLeary:2008myb}:
\begin{align}
b_{\rm max}&\simeq0.21\,{\rm AU}\;\left({m_1m_2\over100M_\odot^2}\right)^{}\left({m_1+m_2\over20M_\odot}\right)^{5\over7}\left({v_{\rm rel}\over{10~{\rm km\ s^{-1}}}}\right)^{-{9\over7}}\,,
\end{align}
where $v_{\rm rel}$ is the relative velocity between the two bodies at infinity (this equation assumes that we are in the gravitational focusing regime).
Considering typical values $m_1=m_2=10M_\odot$ and $v_{\rm rel}=10~\rm km\ s^{-1}$ and using Eq.~(11) of Ref.~\cite{1989ApJ...343..725Q}, we find that the pericenter distance at the closest approach should not exceed $\simeq1.3\times10^{-4}$~AU for the capture to occur.

In our simulations, whenever there is a capture, we sample the impact parameter such that $b^2$ is uniform in $[0, b_{\rm max}^2)$~\cite{1983ApJ...268..319H}.
If $b<b_{\rm max}$, then the amount of GW energy $\delta E_{\rm GW}$ released, given by Eq.~(10) of Ref.~\cite{1989ApJ...343..725Q}, is large enough to extract all of the initial kinetic energy $T$ of the two-body system, leaving behind a binary with internal energy $E'=T-\delta E_{\rm GW}<0$.
Most of the GW energy is radiated during the closest approach and in the direction of the particles' motion, which is why this dissipative phenomenon is also known as ``gravitational bremsstrahlung.''
Gravitational radiation also carries away angular momentum, but the radiated angular momentum is subdominant relative to the initial angular momentum of the system and can be neglected~\cite{OLeary:2008myb}.

Based on the final energy of the system, we calculate the semimajor axis $a$ and eccentricity $e$ of the newly formed bound system to be:
\begin{subequations}
\begin{align}
a&=-{Gm_1m_2\over2E'},\\
e&=\sqrt{1+{2E'b^2v_{\rm rel}^2\over G^2m_1m_2(m_1+m_2)}}.
\end{align}
\end{subequations}

\bibliography{refs}

\end{document}